\documentclass[acmsmall]{acmart}

\usepackage{amsmath}
\usepackage{amssymb}
\usepackage{enumitem}
\usepackage{graphicx}
\usepackage{algorithmic}
\usepackage{amsfonts} 
\usepackage[ruled,vlined]{algorithm2e}
\usepackage{url}
\usepackage{subcaption} 
\usepackage{multirow}
\usepackage{booktabs}
\usepackage{hyperref}
\usepackage{makecell}

\AtBeginDocument{%
  }

\setcopyright{acmlicensed}
\copyrightyear{XXXX}
\acmYear{XXXX}
\acmDOI{XXXXXXX.XXXXXXX}

\acmJournal{JACM}
\acmVolume{XX}
\acmNumber{X}
\acmArticle{XXX}
\acmMonth{8}

\begin{document}

\title{Feature Interaction Fusion Self-Distillation Network For CTR Prediction}

\author{Lei Sang}
\affiliation{%
  \institution{Anhui University}
  \streetaddress{111 Jiulong Rd}
  \city{Hefei}
  \state{Anhui Province}
  \country{China}
  \postcode{230601}
}
\email{sanglei@ahu.edu.cn}

\author{Qiuze Ru}
\affiliation{%
  \institution{Anhui University}
  \streetaddress{111 Jiulong Rd}
  \city{Hefei}
  \state{Anhui Province}
  \country{China}
  \postcode{230601}
}
\email{ruqiuze@stu.ahu.edu.cn}

\author{Honghao Li}
\affiliation{%
  \institution{Anhui University}
  \streetaddress{111 Jiulong Rd}
  \city{Hefei}
  \state{Anhui Province}
  \country{China}
  \postcode{230601}
}
\email{salmon1802li@gmail.com}

\author{Yiwen Zhang} 
\authornote{Corresponding author}
\affiliation{%
  \institution{Anhui University}
  \streetaddress{111 Jiulong Rd}
  \city{Hefei}
  \state{Anhui Province}
  \country{China}
  \postcode{230601}
}
\email{zhangyiwen@ahu.edu.cn}

\author{Qian Cao}
\affiliation{%
  \institution{Chaohu University}
  \streetaddress{No. 1, Bantang Road, Chaohu Economic Development Zone}
  \city{Hefei}
  \state{Anhui Province}
  \country{China}
  \postcode{238000}
}
\email{caoqian@chu.edu.cn}

\author{Xindong Wu}
\affiliation{%
 \institution{Hefei University of Technology}
  \streetaddress{193 Tunxi Road}
  \city{Hefei}
  \state{Anhui Province}
  \country{China}
  \postcode{230601}
 }
\email{xwu@hfut.edu.cn}

\renewcommand{\shortauthors}{Trovato et al.}

\begin{abstract}
 Click-Through Rate (CTR) prediction plays a vital role in recommender systems, online advertising, and search engines.  
Most of the current approaches model feature interactions through stacked or parallel structures, with some employing knowledge distillation for model compression.
However, we observe some limitations with these approaches:
(1) In parallel structure models, the explicit and implicit components are executed independently and simultaneously, which leads to insufficient information sharing within the feature set. (2) The introduction of knowledge distillation technology brings about the problems of complex teacher-student framework design and low knowledge transfer efficiency. (3) The dataset and the process of constructing high-order feature interactions contain significant noise, which limits the model's effectiveness.
To address these limitations, we propose FSDNet, a CTR prediction framework incorporating a plug-and-play fusion self-distillation module. Specifically, FSDNet forms connections between explicit and implicit feature interactions at each layer, enhancing the sharing of information between different features. The deepest fusion layer is then used as the teacher model, utilizing self-distillation to guide the training of shallow layers. Empirical evaluation across four benchmark datasets validates the framework's efficacy and generalization capabilities.
The code is available on \url{ https://anonymous.4open.science/r/FSDNet}.
\end{abstract}




\maketitle

\section{Introduction}
\label{sec:introduction}
Click-Through Rate (CTR) prediction constitutes a fundamental component in recommender systems, online advertising, and search engines  \cite{ openbenchmark, Bars,2024Attacking}. The primary objective of CTR prediction is to predict the probability of user interaction with presented items by analyzing user profiles, item attributes, and context. The significance of accurate CTR prediction is twofold: it enhances user experience through personalized recommendation alignment while optimizing resource allocation in product rankings and advertisement placements   \cite{2021EDCN,2021dcnv2}. Central to achieving high prediction accuracy is the effective modeling of feature interactions  \cite{openbenchmark}. Early approaches heavily relied on expert knowledge and models based on Logistic Regression (LR) \cite{2007LR},  Factorization Machines (FM) \cite{2010FM,2016FFM,2016HOFM},  which were limited to modeling low-order or fixed-order feature interactions. 
 The advent of Deep Neural Networks (DNNs) has revolutionized the field by enabling implicit modeling of complex feature interactions   \cite{2017DCN,2016DNNyoutube}, thereby reducing dependence on manual feature engineering. 
 However, despite their theoretical status as universal function approximators 
  \cite{MLPapproximators}, DNNs exhibit limitations in accurately modeling certain fundamental operations, such as inner products  \cite{neuralvsmf}.

Recent research has shifted focus toward integrating explicit interactions modeled by controlled interaction orders  \cite{2016PNN,2019autoint} to address the limitations of implicit interactions.
CTR models are primarily categorized into parallel  \cite{2017DeepFM,2017DCN,2018xdeepfm,2021dcnv2,2021EDCN} and stacked structures  \cite{2018pnn2,2019FIGNN,2021Xcrossnet,2021masknet}, distinguished by their integration approach. As illustrated in Figure \ref{ensemble},  parallel structure models process explicit and implicit components independently before fusion, while stacked structure models cascade components sequentially.
Since the DNN part in the parallel structure can be seen as a complementary component,  the parallel structure tend to have more stable training compared to the stacked structure \cite{ctr_estimation,2021EDCN}. The pursuit of accuracy has led to sophisticated ensemble structure models incorporating Self-Attention  \cite{2019autoint,2022dexdeepfm}, Mask Mechanisms  \cite{2021masknet}, and Contrastive Learning  \cite{2023Cl4ctr}. However, these advancements introduce a large number of parameters and complex structures,  
 which significantly restricts their application on resource-constrained devices and in real-time tasks.  To address these limitations, knowledge distillation  techniques  \cite{2015KD,2023bkd,2023gkd} enable efficient model compression by transferring knowledge from complex teacher models to compact student architectures. Notable implementations include a multi-teacher ensemble framework with "teacher gating"  \cite{2020KDCTR}, KD-DAGFM's \cite{2023gkd} directed acyclic graph approach for feature interaction distillation, and BKD's  \cite{2023bkd} graph neural network-based methodology for interaction transfer. Despite the success, the aforementioned CTR models still face certain limitations:

\begin{figure}[t]
    \centering
    \begin{subfigure}[b]{0.38\linewidth}
        \centering
        \includegraphics[width=\linewidth]{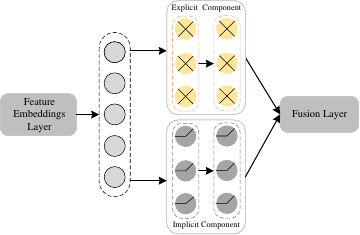}
        \caption{Parallel Structure}
    \end{subfigure}
    \hspace{0.35cm} 
    \begin{subfigure}[b]{0.38\linewidth}
        \centering
        \raisebox{0.6cm}{\includegraphics[width=\linewidth]{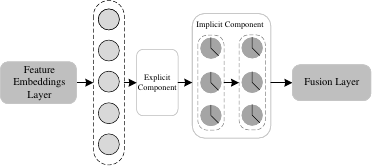}}
        \caption{Stacked Structure}
    \end{subfigure}
    
    \captionsetup{justification=raggedright}
    \caption{The architecture comparison among parallel and stacked structures.}
    \label{ensemble}
    \vspace{-1.1em}
\end{figure}

\textbf{Lack of information sharing between different components.} In parallel structure models, each component processes input data independently \cite{2016widedeep,2017DeepFM,2021dcnv2}, such as DCNv2  \cite{2021dcnv2} independently performs cross network and deep network learning. Since there is no information sharing between hidden layers of different components, they independently learn latent representations and typically only fuse the results at the final layer. This delayed integration paradigm, where components establish isolated representational spaces, inherently constrains the depth and richness of feature information exchange.  Moreover, this architectural isolation may induce gradient bias during backpropagation  \cite{2019Gradientilt}, potentially impeding convergence toward global optima.

\begin{figure}[t]
    \centering
    \begin{subfigure}[b]{0.45\linewidth}
        \centering
     \raisebox{0.25cm}{\includegraphics[width=\linewidth]{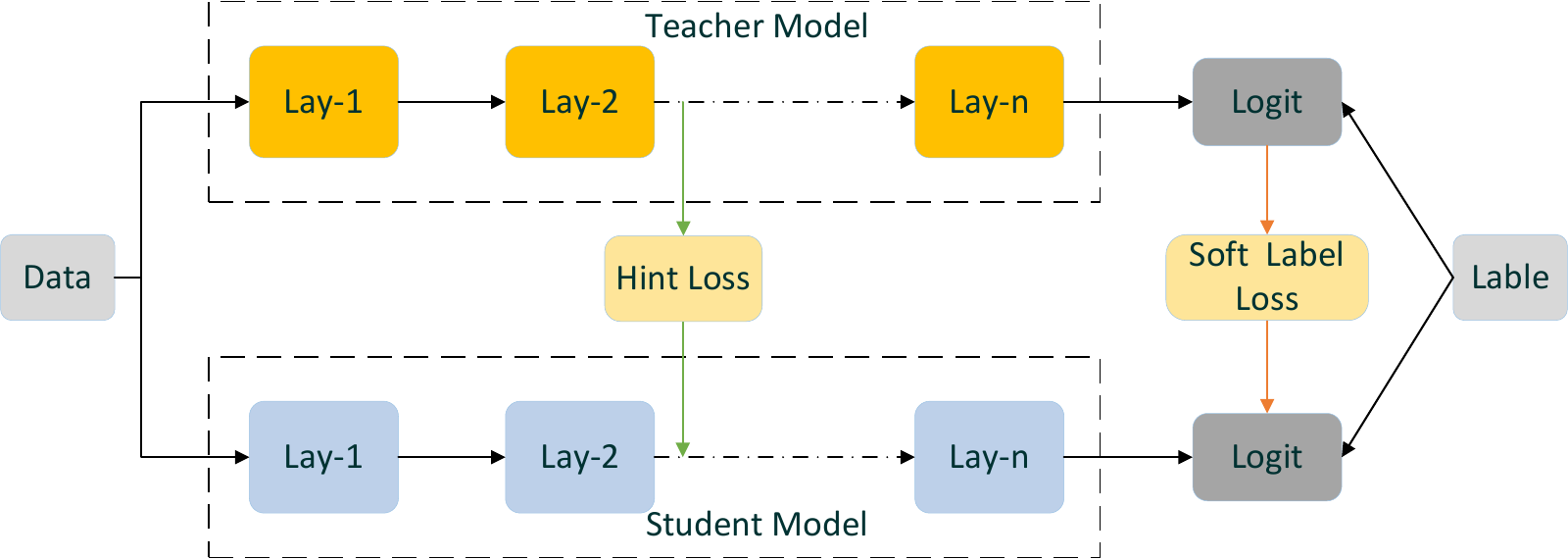}}
        \caption{Knowledge Distillation}
        \label{fig:first_image}
    \end{subfigure}
  \hspace{0.3cm} 
    \begin{subfigure}[b]{0.41\linewidth}
        \centering
        \includegraphics[width=\linewidth]{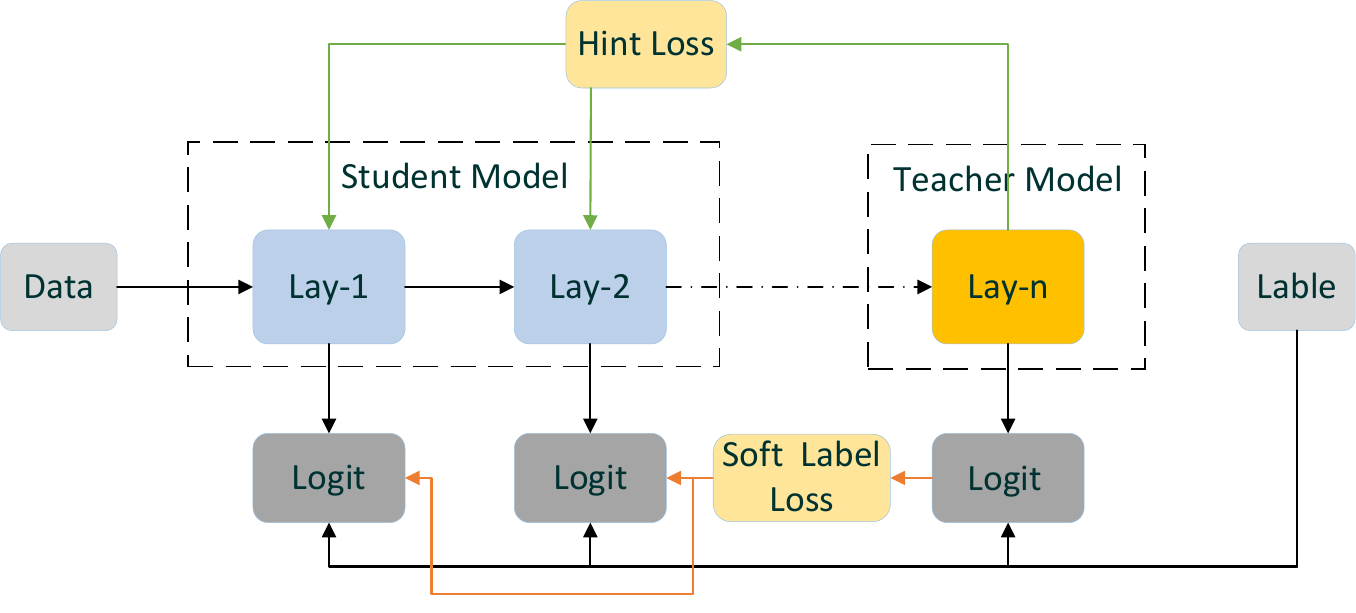}
        \caption{Self distillation}
        \label{fig:second_image}
    \end{subfigure}
    \captionsetup{justification=raggedright}
    \caption{The structural comparison of knowledge distillation and self-distillation.}
    \label{fig:two_images}
    \label{KDVSSD}
     \vspace{-1.1em}

\end{figure}

 \textbf{Complex design and low transfer efficiency in knowledge distillation.}    As depicted in Figure \ref{KDVSSD} (a),  traditional knowledge distillation\cite{2015KD} employs a smaller model (the student model) to imitate the output distributions of a more sophisticated model (the teacher model). 
 However, studies\cite{2020improvedteacherkd,2021knowledgesurvey} indicate  that a model achieving excellent results is not always an effective teacher model.  When multiple teacher models are needed for joint supervision\cite{2020KDCTR}, the teacher-student framework  necessitates intricate design considerations and computational overhead. 
  Furthermore, the compression process of knowledge distillation inherently introduces information degradation \cite{2021knowledgesurvey}, potentially resulting in suboptimal student model performance that may fall short of both the teacher model's accuracy and that of directly trained counterparts.

\textbf{Excessive noise in feature interactions.} 
Users may click on certain items not out of genuine interest, but due to misoperation or other reasons \cite{2021rce,2021noise}. 
These anomalous patterns propagate through the model, leading to the acquisition of spurious feature relationships. Furthermore, the number of feature interactions exhibits linear growth with increasing layer depth. Empirical analyses through hyperparametric studies  \cite{2018xdeepfm,2018pnn2,2023GDCN} demonstrate performance degradation beyond certain interaction orders, particularly exceeding third order. This phenomenon suggests the emergence of ineffective interactions, potentially amplifying the noise propagation throughout the learning process.

In this paper, we propose a fusion self-distillation module to address the prevalent limitations discussed above.  This is a lightweight, plug-and-play module that consists of two parts: feature interaction fusion and self-distillation. It can be broadly applicable to various parallel structured models.  In addition, we propose a new CTR parallel structure framework  based on DCNv2, called the Feature Interaction Fusion Self-Distillation Network (FSDNet).
First,  we establish connections between each layer outputs of the cross network and the deep network to resolve the problem of insufficient information sharing in the parallel structure. Then, we introduce self-distillation, which is grounded in knowledge distillation \cite{2021knowledgesurvey}. 
 As shown in Figure \ref{KDVSSD} (b), self-distillation does not use an additional teacher model. Instead, the deepest network is used as the teacher model.
 Specifically, we generate predictions by passing the fused feature information from each layer through a linear activation layer.  The deepest fusion layer is regarded as the teacher model, with soft label loss and hint loss used to guide each of the previous shallower layer (i.e., the student models). This process helps stabilize the parallel structure and improve performance. In addition, the smoother soft labels contribute to the model's generalization ability and robustness to noise.

In summary, the key contributions of this paper are as follows:
\begin{itemize}
    \item We analyze three limitations of current parallel structure models: insufficient information sharing, low transfer efficiency in knowledge distillation, and excessive noise in feature interactions.
To address these limitations, we propose a novel CTR framework, FSDNet.
    \item  We introduce a plug-and-play fusion self-distillation module composed of two parts: information fusion and self-distillation. The former establishes  connections between explicit and implicit feature interactions, while the latter improves the framework's accuracy and robustness via  self-hinting.
    
    \item We conduct extensive experiments across four datasets, demonstrating the effectiveness and robustness  of the proposed FSDNet framework. Additionally, we validate the compatibility of the fusion self-distillation module across various parallel structure CTR models.
    
\end{itemize}

\section{Preliminaries}
\subsection{CTR Prediction Task}
The click-through rate prediction task aims to predict the likelihood that a user will click on a piece of content, such as an advertisement or recommendation.
CTR prediction is typically formulated as a binary classification problem \cite{2018pnn2,2021masknet}. Given a record  containing  a user $\mathbf{u}$ and an item $\mathbf{i}$ (including user profiles, item attributes, and context), the model outputs a prediction between 0 and 1,  representing the probability  of a click. These three types of features generally consist of the following:
\begin{itemize}
    \item  \emph{User profiles ($\mathbf{x_p} $):} age, gender, interest, occupation, etc.
    \item  \emph{Item attributes ($\mathbf{x_a}$):} category, title, price, etc.
    \item \emph{Context ($\mathbf{x_c}$):} time, device type, weather, location, etc.
\end{itemize}
We use a tuple data format to define samples in CTR prediction, where each record represents a  behavior: $\mathbf{X}=\{\mathbf{x_p},\mathbf{x_a},\mathbf{x_c}\}$.  $y \in \{0, 1\} $ represents the true label of the user’s click behavior. When y=1, it indicates that the user clicked on this item, which is considered a positive sample. When y=0, it is considered a negative sample. The primary objective of CTR prediction is to construct a model based on historical data to perform probability predictions. In recent CTR models, feature interaction modeling has become the key component to improving predictive performance.  By using explicit or implicit methods, the model learns relationships between features at different orders. A typical framework for a basic CTR model is as follows:
\begin{equation}
    \hat{y}=\mathrm{Model}(\mathbf{u},\mathbf{i},g\{\mathbf{e}_1,\mathbf{e}_2,\ldots,\mathbf{e}_f\};\theta),
\end{equation}
where $\{\mathbf{e}_1,\mathbf{e}_2,\ldots,\mathbf{e}_f\}$ represents features processed into low-dimensional dense vectors, $\theta$ represents the parameters of the model, and $g(\cdot)$ represents the function of constructing interactions.

CTR prediction usually optimizes the model by reducing the discrepancy between the predicted click rate and the actual click situation: $\min\lVert y-\hat{y}\rVert.$

\subsection{Deep \& Cross Network v2}
The  DCNv2 \cite{2021dcnv2} is an improved model compared to DCN, which is representative and efficient. It consists of two modules: the Cross Network 
and the Deep Network: $\phi_{\text{ DCN-v2}}(\mathbf{x}) = \phi_{\text{Cross}}(\mathbf{x}) + \phi_{\text{DNN}}(\mathbf{x}).$
Specifically, module $\phi_{\text{Cross}}(\mathbf{x})$ gradually generates higher-order feature interactions through computations at each layer to capture explicit feature interactions. And module $\phi_{\text{DNN}}(\mathbf{x})$ automatically learns complex and implicit feature interactions in the hidden layers using the nonlinear activation functions of deep neural networks. 
To facilitate understanding, we integrate the proposed fusion self-distillation module into this parallel structure, which significantly improves its performance.

\begin{figure*}[t]
  \centering
  \includegraphics[width=\textwidth]{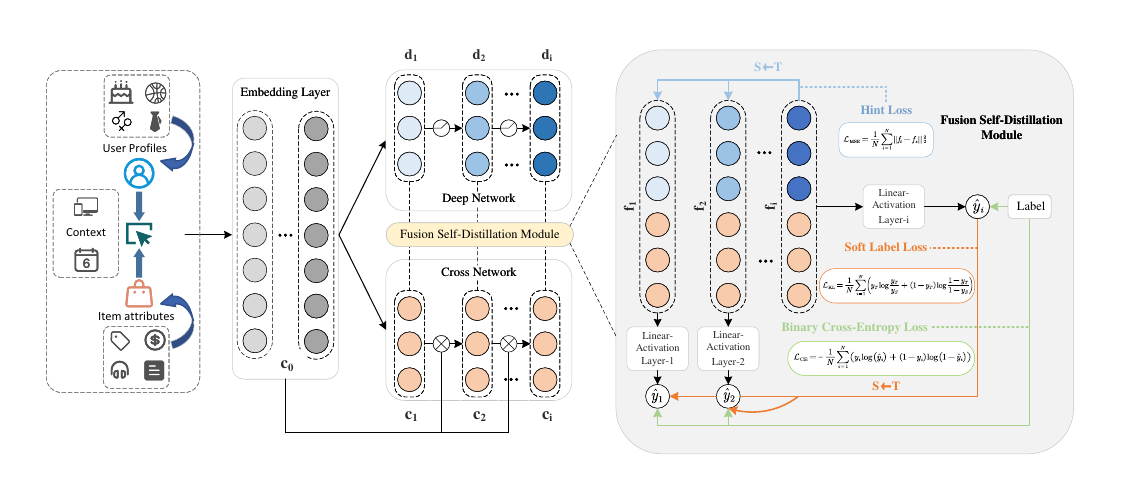}
  \captionsetup{justification=raggedright}
  \caption{The overall framework of FSDNet. The embedding layer maps sparse vectors to low-dimensional dense embeddings, while the cross network captures explicit feature interactions and the deep network models implicit feature relationships. The fusion self-distillation module enhances the framework's performance and generalization ability by connecting the outputs from both networks and using the predictions from the final layer to guide the learning of earlier layers. The linear-activation layer represents the simultaneous use of a linear layer and an activation function to generate prediction values, and $\otimes$ represents the cross operation in Eq. \ref{crossoperation}.} 
  \label{fig:FSDNet}
\end{figure*}

\section{FSDNet: Feature Interaction Fusion Self-Distillation
Network}
In this section, we present a novel  framework FSDNet in detail. The  framework is shown in Figure  \ref{fig:FSDNet}. The framework comprises four primary components: an embedding layer, a cross network, a deep network, and a fusion self-distillation module. The embedding layer transforms sparse input vectors into dense, lower-dimensional representations. Subsequently, the cross network is responsible for modeling explicit feature interactions, while the deep network focuses on capturing implicit, higher-order interactions. The fusion self-distillation module enhances the network's performance by concatenating the outputs of corresponding layers from both the deep and cross networks. This module then leverages the predictions from the final layer as a teacher model, guiding the learning process of preceding layers through knowledge distillation.  This fusion self-distillation module possesses excellent generalization capabilities and is broadly applicable to parallel structural models in the CTR domain.

\subsection{Embedding Layer}
\label{EmbeddingLayer}
For CTR prediction, input features can be categorical, numerical, or multi-valued types (e.g.,  several interest tags for a user). By utilizing the techniques of one-hot encoding \cite{openbenchmark,2019FIGNN}, these input samples can be transformed into vector. However, due to the high sparsity of many features, this leads to high-dimensional vector spaces. The embedding layer is typically applied to map these vectors to dense, low-dimensional embeddings:
\begin{equation}
    \mathbf{e}_i = E_i x_i,
\end{equation}
where $x_i \in X$ denotes a vector of $ i^{th} $ field in the sample $X$, and $E_i \in \mathbb{R}^{v_i \times d}$ represents the embedding matrix. Here, $v_i$ refers to the vocabulary size, and $d$ stands for the embedding dimension. Afterwards, the resulting features are  concatenated to yield the final embedding layer representation as:
 \begin{equation}
 \mathbf{h}=concat\left(\mathbf{e}_1, \mathbf{e}_2, \cdots, \mathbf{e}_f\right),
 \end{equation}
where $f$ represents the count of fields.

Unlike many existing methods \cite{2017DeepFM,2018pnn2,2019autoint} that require uniform embedding sizes for specific interaction operations, our model supports arbitrary embedding dimensions, which is especially beneficial for industrial recommender systems.

\subsection{Feature Interaction Construction}
\label{CrossNetwork}
To construct arbitrary finite-order explicit interactions between different features, we employ the cross network optimized by DCNv2 \cite{2021dcnv2}. The $(l+1)^{th}$  layer is formulated as follows:
\begin{equation}
\mathbf{c}_{l+1} = \mathbf{c}_0 \odot (\mathbf{W}^c_l \mathbf{c}_l + \mathbf{b}_l^c) + \mathbf{c}_l,
\label{crossoperation}
\end{equation}
where $\mathbf{c}_0 \in \mathbb{R}^d$ represents the embedding layer output, containing the original first-order features. The outputs from the $l^{th}$ and $(l+1)^{th}$ layers are denoted by $\mathbf{c}_l$ and $\mathbf{c}_{l+1}$, while $ W^c_l $ and $ b_l^c $ indicate the learnable weight matrix and bias vector.  In an $l$-layer cross network, the maximum expressible polynomial order it can represent is $l+1$, capturing all feature interactions up to that order.  Additionally, the cross network can be viewed as performing both bit-wise and feature-wise interactions simultaneously.
\label{DeepNetwork}

The deep network is responsible for building implicit feature interactions.  By learning more profound levels of feature abstraction, it explores complex patterns and dependencies that the cross network may fail to detect. The $(l+1)^{th}$ layer in the deep network can be represented using the following formula:
\begin{equation}
    \mathbf{d}_{l+1}=f(\mathbf{W}_l^d\mathbf{d}_l+\mathbf{b}_l^d),
\end{equation}
where $\mathbf{d}_{l}$ , $\mathbf{d}_{l+1}$ denote the outputs from the $(l+1)^{th}$ and  $l^{th}$ layers. 
The weight matrix is $W_l^d\in \mathbb{R}^{n_{l+1} \times n_l}$,  and the bias vector is $\mathbf{b}_l^d \in \mathbb{R}^{n+1}$. $f(\cdot)$ denotes the activation function, with ReLU being utilized in this context. It is crucial to note that both the cross network and the deep network share the same initial input, denoted as: $\mathbf{c}_0=\mathbf{d}_0=\mathbf{h}$.

\subsection{Fusion Self-Distillation Module}
\label{FusionSelf-Distillation}
 We incorporate a self-distillation technique from knowledge distillation  \cite{2019Beyourownteacher,2021knowledgesurvey}, which is a lightweight and self-hinting approach. First, we associate the output of each layer from  the cross network and the deep  network to form  new layers. The output fused feature information is represented as  $\mathbf{f}_l=f(\mathbf{c}_l, \mathbf{d}_l)$.
 During the training phase, the deepest fused layer serves as the teacher model, utilizing distillation to instruct the preceding, shallower sections (i.e., the student models). This strategy leverages the higher-level knowledge contained in the deeper layers to enhance the training of the earlier layers, eliminating the need for the external teacher model. This facilitates a more effective and consolidated learning process within the model itself.  Moreover, during backpropagation, self-distillation can optimize the interaction between each layer of the cross and deep networks through gradients. 
 
To transfer knowledge from the teacher model to the student model, we introduce three distinct types of loss functions during training. Each loss function is specifically crafted to optimize different aspects of the student model's learning process.
\subsubsection{Binary Cross-Entropy Loss}
The fused information $\mathbf{f}_l$ from each connection layer is processed through a linear-activation layer to produce the prediction value:
$\hat{y}^l=\sigma(\mathbf{w}^\top\mathbf{f}_l+\mathbf{b})$, where $\sigma$ is the $Sigmoid$ activation function. Next, the true labels  $y$ from the training dataset are used to guide the prediction results from the deepest layer to all shallow layers:
\begin{equation}
\label{Binaryloss}
\mathcal{L}_{\text{CE}}^l=-\frac{1}{N}\sum_{i=1}^N{\left( y_i\log \left( \hat{y}_i^l \right) +\left( 1-y_i \right) \log \left( 1-\hat{y}_i^l \right) \right)},
\end{equation}
where $\mathcal{L}_{\text{CE}}^l$ indicates the binary cross-entropy loss at layer $l$, a loss function commonly applied in CTR models. $N$ refers to the total count of training samples. This method allows the hidden knowledge within the dataset to be directly infused into every layer of the framework via the labels. The total cross-entropy loss, denoted as $\mathcal{L}_{\text{CE}}$, is calculated as the sum of losses from each layer: $\mathcal{L}_{\text{CE}}=\mathcal{L}_{\text{CE}}^1+...+\mathcal{L}_{\text{CE}}^l+...+\mathcal{L}_{\text{CE}}^n$
, where $n$ indicates the total count of layers. The framework's final prediction value is derived by averaging the prediction outcomes of all layers:
\begin{equation}
    \hat{y}=\frac{1}{n}\sum_{i=1}^ny_i^l.
\end{equation}
\subsubsection{Soft Label Loss}
Soft labels \cite{2015KD} are obtained by applying temperature scaling to the teacher model's probability output: $y_T= \sigma(\frac{\mathbf{z}_T}{\tau})$, where $\mathbf{z}_T$ is the logit value obtained by the teacher model through the linear layer, and $\tau$ represents the distillation temperature, which produces softer label probabilities. The student model's output is also temperature scaled: $y_S= \sigma(\frac{\mathbf{z}_S}{\tau})$.
 We apply the KL (Kullback-Leibler) Divergence Loss to penalize discrepancies between the outputs of the student and teacher models:
\begin{equation}
\label{SoftLabel}
\mathcal{L}_{\text{KL}}^\text{T, S}=\frac{1}{N}\sum_{i=1}^N{\left( y_T\log \frac{y_T}{y_S}+\left( 1-y_T \right) \log \frac{1-y_T}{1-y_S} \right)}.
\end{equation}
   In this loss, the student model improves performance by matching the teacher model’s soft labels. At the same time, we do not isolate  the output of the teacher model, allowing it to adaptively adjust its embeddings based on the loss to better provide useful knowledge signals to the student network.
Compared with the true labels, soft labels provide smoother decision boundaries, enabling the student model to generalize more effectively and reducing the impact of incorrect labels on the model. The loss promotes information sharing between different layers, and this collaborative learning enhances the extraction of effective feature interactions, thereby reducing the irrelevant noise introduced by high-order interactions.
 The ultimate soft label loss is the sum of each pair of student and teacher models: $\mathcal{L}_{\text{KL}}=\mathcal{L}_{\text{KL}}^\text{T, 1}+...+\mathcal{L}_{\text{KL}}^\text{T, S}+...+\mathcal{L}_{\text{KL}}^\text{T, N}$, where $N$ refers to the number of shallow models in total.

\subsubsection{Hint Loss} 
Hint loss provides fine-grained supervision signals, aimed at guiding the student  to learn the implicit knowledge in the teacher model's feature maps \cite{2014fitnets}. This additional supervision signal improves training efficiency and performance, helping the model consolidate important feature representations learned during the early stages of training. Its mechanism works by minimizing the distance between shallow feature maps and the deepest feature maps within the framework, and we do not limit the impact of the loss on the teacher model. We achieve this using the Mean Squared Error (MSE) Loss:
\begin{equation}
\label{HintLoss}
\mathcal{L}_{\text{MSE}}^\text{T, S}=\frac{1}{N}\sum_{i=1}^N{|}|\mathbf{f}_T-\mathbf{f}_S||_{2}^{2},
\end{equation}
where $\mathbf{f}_T$ and $\mathbf{f}_S$ represent the embedding of the teacher model and the student model, respectively. When each student-teacher model pair is combined, the final hint loss function becomes: $\mathcal{L}_{\text{MSE}}=\mathcal{L}_{\text{MSE}}^\text{T, 1}+...+\mathcal{L}_{\text{MSE}}^\text{T, S}+...+\mathcal{L}_{\text{MSE}}^\text{T, N}$, where $N$ represents the total count of shallow models.

To summarize, we employ binary cross-entropy loss, soft label loss, and hint loss to jointly optimize the FSDNet model. The overall loss function can be expressed as:
\begin{equation}
\label{totalloss}
\mathcal{L}=\mu\mathcal{L}_{\text{CE}}+(1-\mu)\mathcal{L}_{\text{KL}}+\gamma\mathcal{L}_{\text{MSE}},
\end{equation}
where $\mu$ and $\gamma$ are hyperparameters that regulate the balance between the different loss functions.  To clearly illustrate the entire training process of the FSDNet framework, a complete workflow is shown in Algorithm \ref{Algorithm}.

\subsubsection{Deep and Cross Layer Combination} 
We concatenate the outputs of each layer from the cross and deep networks  to form  new fusion layers.
By using the self-distillation method, the framework optimizes the two networks through gradients during the backpropagation process. This approach facilitates the integration of explicit and implicit feature interactions, allowing the framework to utilize the characteristics of both feature types more effectively at each layer, thereby addressing the limitation  of inadequate information sharing in  parallel structure models.  The impact of different connection methods is examined in Section \ref{Combination}.  Based on the experimental results, we conclude that directly concatenating  the outputs of each layer from the cross and deep networks yields the best performance without introducing additional parameters. The concatenated layer is represented as $\mathbf{f}_l=[\mathbf{c}_l; \mathbf{d}_l]$.

\begin{algorithm}[t]
\footnotesize
\caption{The training process of FSDNet\label{training_process}}
  \SetAlgoLined 
  \KwIn{input samples $X \in N$;}
  \KwOut{model parameters $\Theta$; }
  Initialize parameters $\Theta$\;
  \While{FSDNet has not reached the patience threshold for early stopping}{
   \For{$X \in N$}{
      $\mathbf{h}$ ← Embedding Layer $\leftarrow$ $X$ according to Section \ref{EmbeddingLayer}\; 
      Get $\mathbf{c}_l$ and  $\mathbf{d}_l$
      from the explicit and implicit components according to Section \ref{CrossNetwork}\;
      $\hat{y}$ $\leftarrow$ $\hat{y}^l$ $\leftarrow$  $\mathbf{f}_l$ $\leftarrow$ 
      $ f(\mathbf{c}_l, \mathbf{d}_l)$ according to Section \ref{FusionSelf-Distillation}\;
      Calculate     
      $\mathcal{L}_{\text{CE}} $ $\leftarrow$
     $\mathcal{L}^l_{\text{CE}}$,  $\mathcal{L}^l_{\text{CE}}$
      according to Eq.(\ref{Binaryloss})\;
      $\hat{y}_T$ guides $\hat{y}_S$ by calculating Soft Label Loss $\mathcal{L}_{\text{KL}}$ according to Eq.\ref{SoftLabel}\; 
      $f_T$ guides $f_S$ by calculating Hint Loss $\mathcal{L}_{\text{MSE}}$ according to Eq.\ref{HintLoss};
      
     Get total loss according to Eq. (\ref{totalloss})\;

   }
   Get Average gradients from mini-batch\;
   Update parameters by decreasing the gradients $\mu\nabla_{\Theta}\mathcal{L}_{\text{CE}} + (1-\mu)\nabla_{\Theta}\mathcal{L}_{\text{KL}}+\gamma\nabla_{\Theta}\mathcal{L}_{\text{MSE}}$;
    }
    \Return model parameters $\Theta$;
\label{Algorithm}
\end{algorithm}

\subsection{Gradients Analysis}
In the CTR parallel structure models, some studies \cite{2019autoint,2018xdeepfm} use the logical sum fusion method  for different components: $\hat{y} = \sigma(z_c+z_d)$, while we use the method of connecting the vectors output by different networks \cite{2017DCN,2023GDCN}, the logit value: 
$z=\mathbf{w}^\top[\mathbf{c},\mathbf{d}]+\mathbf{b}$, the prediction value:
$\hat{y} = \sigma(z)$, which usually contains more information. In FSDNet, the outputs from each layer of the cross and the deep network are concatenated together, and this  combined output is fed into the subsequent linear-activation layer for prediction.
The framework’s loss function consists of three components: binary cross-entropy loss, soft label loss, and hint loss.  These three parts of loss will work together to update the parameters of the model. Next, we will analyze the gradient transfer during the training process.

\subsubsection{Gradients of $\mathcal{L}_{\text{CE}}$ Loss}
The logit values from each layer of the framework are transformed into prediction values through the sigmoid function, expressed as:
\begin{equation}
    \hat{y}=\sigma(z)=\frac1{1+e^{-z}}.
\end{equation}
According to Eq. \ref{Binaryloss}, the gradient of $\mathcal{L}_{\text{CE}}$ for $\hat{y}$ can be derived as:
\begin{equation}
    \frac{\partial \mathcal{L}_{\text{CE}}}{\partial\hat{y}}=-\frac y{\hat{y}}+\frac{1-y}{1-\hat{y}}.
\end{equation}
The gradient of the loss $\mathcal{L}_{\text{CE}}$ regarding the logit value $z$ is:

\begin{equation}
  \frac{\partial\mathcal{L}_{\text{CE}}}{\partial z} = \frac{\partial\mathcal{L}_{\text{CE}}}{\partial\hat{y}}\cdot\frac{\partial\hat{y}}{\partial z} = \left(-\frac{y}{\hat y} + \frac{1-y}{1-\hat y}\right) \cdot \hat y(1-\hat y) = \hat{y} - y = \sigma(z) - y.
\end{equation}
The gradients of the loss $\mathcal{L}_{\text{CE}}$ relative to the output of the cross network and the deep network are:
\begin{equation}
    \frac{\partial \mathcal{L}_{\text{CE}}}{\partial \mathrm{\mathbf{c}}}=(\hat{y}-y)\cdot W^\mathrm{c},  
    \frac{\partial \mathcal{L}_{\text{CE}}}{\partial \mathrm{\mathbf{d}}}=(\hat{y}-y)\cdot W^\mathrm{d}.
\end{equation}
where $\mathbf{W^c}$ and $\mathbf{W^d}$ are the corresponding sub-matrices of $W$ respectively. These formulas show that the deviation from the prediction value is simultaneously propagated to both parts of the network.  In the linear-activation layer of each layer, the outputs of the cross network and the deep network are mixed together and jointly optimized through the propagation of gradients.
\subsubsection{Gradients of $\mathcal{L}_{\text{KL}}$ Loss}
Referring to Eq. \ref{SoftLabel}, the gradient of KL divergence loss for  the student model’s logit value $z_S$  is:
\begin{equation}
\frac{\partial\mathcal{L}_{\text{KL}}}{\partial z_S}=\left(-\frac{\hat{y}_T}{\hat{y}_S}+\frac{1-\hat{y}_T}{1-\hat{y}_S}\right)\cdot\sigma'(z_S).
\end{equation}
When the prediction value of the student model deviates greatly from the teacher model’s output, the absolute value of the gradient will be larger, resulting in a larger update of the model parameters. This process brings the probability distribution of the student model closer to that of the teacher model.

\subsubsection{Gradients of $\mathcal{L}_{\text{MSE}}$ Loss}
Combined with Eq. \ref{HintLoss}, the gradient of MSE loss to the output embedding $\mathbf{f}_S$ of the student model is:
\begin{equation}
    \frac{\partial L_\mathrm{MSE}}{\partial\mathbf{f}_S}=-(\mathbf{f}_T-\mathbf{f}_S).
\end{equation}
The meaning of this gradient is very intuitive. It ensures that the student model is always reducing the gap with the teacher model embedding during the learning process and gradually approaching the feature representation of the teacher model.

\section{EXPERIMENTS}
In this section, we perform extensive experiments using four real-world datasets. These experiments aim to evaluate  the effectiveness of our proposed FSDNet framework, as well as the compatibility and robustness of the self-distillation module.  Specifically, we focus on addressing the following research questions (RQs):
\begin{itemize}[leftmargin=*]
\item \textbf{RQ1} How does FSDNet perform in CTR prediction scenarios compared to existing methods?  Is it effective when handling large-scale, highly sparse data?
\item \textbf{RQ2} Can the proposed fusion self-distillation module be generalized to different models and improve performance?
\item \textbf{RQ3} How should the hyperparameters of FSDNet be configured?
\item \textbf{RQ4} How does the model performance change when different key components of the fusion self-distillation module are removed individually?
\item \textbf{RQ5} How do different methods of constructing explicit and implicit feature interaction connections in FSDNet affect model performance?
\item \textbf{RQ6} What effect does the fusion self-distillation module have on feature representations at each layer?
\item \textbf{RQ7} Does the fusion self-distillation module improve FSDNet's robustness to noise?

\end{itemize}

\begin{table}[t]
\small
\renewcommand\arraystretch{1}
\centering
\caption{Dataset statistics.}
\label{dataset}
\resizebox{0.6\linewidth}{!}{
\begin{tabular}{ccccc} 
\toprule 
\textbf{Dataset} & \textbf{\#Instances} & \textbf{\#Fields} & \textbf{\#Features} & \textbf{\#Split} \\
\midrule 
\textbf{Criteo}  & 45,840,617  & 39 & 5,549,252 & 8:1:1\\
\textbf{ML-tag} & 2,006,859  & 3  & 88,596 & 7:2:1\\
\textbf{ML-1M} & 1,006,209  & 5  & 9,626 & 8:1:1\\
\textbf{Frappe} & 288,609 & 10  & 5,382 & 7:2:1\\
\bottomrule
\end{tabular}}
\end{table}

\subsection{Experimental Settings}

\textbf{Datasets.}
We choose four commonly used real-world datasets to evaluate FSDNet against other CTR models: Criteo\footnote{\url{https://www.kaggle.com/c/criteo-display-ad-challenge}}  \cite{openbenchmark}, ML-tag\footnote{\url{https://github.com/reczoo/Datasets/tree/main/MovieLens}}  \cite{AFN}  \cite{Bars}, ML-1M\footnote{\url{https://grouplens.org/datasets/movielens}}  \cite{2019autoint}, and Frappe\footnote{\url{http://baltrunas.info/research-menu/frappe}}  \cite{AFN, frappe}.
Table \ref{dataset} contains a brief introduction to these datasets.  More comprehensive descriptions can be found in the provided references and links.

To maintain fairness,  we follow the protocol established in  \cite{openbenchmark} for data preprocessing. Specifically, we set a threshold, and uncommon features that appear smaller than this threshold will be replaced with a default  "OOV" token.  Given the Criteo dataset's vast size and extreme sparsity (over 99.99\%)  \cite{2019autoint}, we set the threshold to 10. For other datasets, a lower threshold of 2 is used.
Additionally, for the Criteo dataset, we discretize numerical features by applying the following transformation: each numeric value $x$ is rounded down to $\lfloor \log^2(x) \rfloor$ for $x$ greater than 2, and set to 1 otherwise\footnote{\url{https://www.csie.ntu.edu.tw/~r01922136/kaggle-2014-criteo.pdf}}.

\noindent\textbf{Evaluation Metrics.}
For performance evaluation across all models, we use two commonly adopted metrics: AUC and Logloss. AUC represents the area under the ROC curve, measuring the probability that a positive instance ranks above a randomly selected negative instance, assessing the model's overall ranking ability. 
 Logloss, calculated as binary cross-entropy, reflects the precision of the model's predicted probability. It is important to note that, in the CTR prediction task, researchers generally believe that the AUC increase ($\uparrow$) or Logloss decrease ($\downarrow$) of  \textit{\textbf{0.1\%-level}} represents a significant improvement, which has been emphasized many times in previous studies.
 
\noindent\textbf{Baselines.} To illustrate  the effectiveness of  FSDNet, we perform  comparative analyses against several state-of-the-art(SOTA) models:
\begin{itemize}
\item \textbf{FM \cite{2010FM}}: It captures second-order interactions between features by decomposing latent vectors.
\item \textbf{DNN \cite{2016DNNyoutube}}: It extracts higher-order features from data through multiple layers of nonlinear transformations.
\item  \textbf{PNN \cite{2016PNN}}: It introduces a product layer within deep neural networks to  model feature interactions automatically.
\item \textbf{Wide\ \&\ Deep \cite{2016widedeep}}: This model merges a linear model with a feed-forward neural network to strike a balance between memorization and generalization.
\item \textbf{DeepFM \cite{2017DeepFM}}: It combines FM and DNN in parallel while sharing embeddings, allowing it to capture both low-order and high-order feature interactions simultaneously.
\item \textbf{DCN \cite{2017DCN}}: This model proposes a Cross Network to explicitly model feature interactions, combining it with a deep network in parallel.
\item \textbf{xDeepFM \cite{2018xdeepfm}}: This model, an improvement on DeepFM, introduces Compressed Interaction Network (CIN) to model explicit feature interactions at the vector-wise level.
\item \textbf{FiGNN \cite{2019FIGNN}}: It utilizes a fully connected graph to model features and leverages a gated graph neural network to construct high-order interactions between them.
\item \textbf{AutoInt+ \cite{2019autoint}}: It uses a multi-head attention mechanism to automatically capture arbitrary-order interactions between features.
\item \textbf{AFN+ \cite{AFN}}: This model incorporates a logarithmic transformation layer to adaptively learn interactions of arbitrary order among features.
\item \textbf{DCNv2 \cite{2021dcnv2}}: This model introduces a more complex Cross Network structure on the basis of the original DCN, further enhancing its capacity to capture feature interactions.
\item \textbf{EDCN \cite{2021EDCN}}: It proposes a bridge module to capture the interactive signals between different components within the parallel structure and selects features for each hidden layer through a regulation module.
\item \textbf{MaskNet \cite{2021masknet}}: This model proposes MaskBlock as the basic building block and introduces feature-wise multiplication via instance-guided mask.
\item \textbf{GraphFM \cite{2021GraphFM}}: This model utilizes a graph-oriented approachs and models feature interactions through Interaction Selection and Interaction Aggregation.
\item \textbf{CL4CTR \cite{2023Cl4ctr}}: It proposes a CTR contrastive learning framework  that aims to produce high-quality feature representations using a self-supervised approach.

\end{itemize}

\noindent\textbf{Implementation Details.}
All models are developed using PyTorch, based on existing research, and optimized using the Adam optimizer. The embedding dimension is set to 16, and the learning rate is initialized at 0.001. 
During training, we apply the Reduce-LR-on-Plateau scheduler, which reduces the learning rate by a factor of ten if there is no improvement in the monitored metric.
 To avoid  overfitting, early stopping is employed with a patience setting of 2. For fair comparison, the MLP hidden units are set to [400, 400, 400]. The batch size is configured as 4096 for the Criteo dataset and 10,000 for other datasets.
We refer to  \cite{Bars,openbenchmark}, and the original papers of the baseline models to set and fine-tune the hyperparameters.

\subsection{Overall Performance (RQ1)}
We evaluate FSDNet against 15 selected baseline models, with the overall performance summarized in Table \ref{baseline}. Our observations are as follows:
\begin{itemize}
\item Compared to other high-order baseline models, the second-order feature interaction model, FM, exhibits poorer performance. This underscores the importance of high-order modeling to capture complex feature interactions in the CTR domain.
\item Ensemble methods have demonstrated competitive performance across various datasets by integrating explicit and implicit feature interactions, categorized into two types:  stacked and parallel structures. Models based on the stacked structure (e.g., PNN, FiGNN, MaskNet) enhance expressive  power by sequentially chaining two components. Conversely, models based on the parallel structure (Wide\ \&\ Deep, DCN, AutoInt+)  share the same embedding layer among components to  construct explicit and implicit interactions concurrently. These results underscore the importance of modeling both feature types simultaneously.

\begin{table*}[t]
\centering
\renewcommand\arraystretch{1.3}
\caption{Performance comparison across four datasets. Note that the baseline with a “+” suffix indicates the version integrated with a DNN network. Results are presented with the best values in bold and the second-best underlined.  Additionally, we performed a two-tailed T-test ($p$-value) to statistically evaluate FSDNet against the best baseline, with results showing 
$p < 0.01$.  In CTR research, an improvement of \textit{\textbf{0.1\%}} in Logloss and AUC is generally regarded as statistically significant   \cite{2017DCN,2021EDCN,2023Cl4ctr,openbenchmark}.} 
\resizebox{0.9\textwidth }{!}{
\begin{tabular}{c|c|cc|cc|cc|cc}
\Xhline{1px}
\hline
\multirow{2}{*}{Year} & \multirow{2}{*}{Model} & \multicolumn{2}{c|}{Criteo} & \multicolumn{2}{c|}{ML-tag} & \multicolumn{2}{c|}{ML-1M} & \multicolumn{2}{c}{Frappe} \\ \cline{3-10} 
     &                  & AUC $\uparrow$   & Logloss $\downarrow$ & AUC $\uparrow$   & Logloss $\downarrow$ & AUC $\uparrow$   & Logloss $\downarrow$ & AUC  $\uparrow$  & Logloss $\downarrow$ \\ \hline
2010 & FM \cite{2010FM}               & 0.8076 & 0.4443  & 0.9425 & 0.2775  & 0.7920 & 0.5409  & 0.9672 & 0.2029  \\
2016 & DNN \cite{2016DNNyoutube}             & 0.8128 & 0.4393  & 0.9682 & 0.2125  & 0.8116 & 0.5206  & 0.9811 & 0.1653  \\
2016 & PNN \cite{2016PNN}             & 0.8138 & 0.4380  & 0.9691 & 0.2092  & 0.8124 & 0.5179  & 0.9828 & 0.1556  \\
2016 & Wide \&Deep \cite{2016widedeep}     & 0.8135 & 0.4382  & 0.9692 & 0.2105  & 0.8131 & 0.5183  & 0.9832 & 0.1525  \\
2017 & DeepFM  \cite{2017DeepFM}         & 0.8139 & 0.4380  & 0.9694 & \underline{0.2077}  & 0.8139 & 0.5204  & 0.9837 & 0.1575  \\
2017 & DCN   \cite{2017DCN}           & 0.8138 & 0.4383  & \underline{0.9702} & 0.2238  & 0.8143 & 0.5162  & 0.9838 & 0.1544  \\
2018 & xDeepFM  \cite{2018xdeepfm}        & \underline{0.8140} & 0.4382  & 0.9692 & 0.2110  & 0.8142 & \underline{0.5149}  & 0.9844 & \underline{0.1454}  \\
2019 & FiGNN  \cite{2019FIGNN}          & 0.8124 & 0.4395  & 0.9510 & 0.2605  & 0.8108 & 0.5183  & 0.9648 & 0.2266  \\
2019 & AutoInt+  \cite{2019autoint}       & \underline{0.8140} & \underline{0.4378}  & 0.9698 & 0.2266  & 0.8138 & 0.5182  & 0.9841 & 0.1520  \\
2020 & AFN+  \cite{AFN}           & 0.8130 & 0.4392  & 0.9609 & 0.2666  & 0.8141 & 0.5172  & 0.9819 & 0.1598  \\
2021 & DCNv2 \cite{2021dcnv2}           & 0.8139 & 0.4385  & 0.9701 & 0.2254  & \underline{0.8145} & 0.5200  & \underline{0.9845} & 0.1581  \\
2021 & EDCN \cite{2021EDCN}            & 0.8136 & 0.4386  & 0.9603 & 0.2649  & 0.8133 & 0.5178  & 0.9841 & 0.1620  \\
2021 & MaskNet  \cite{2021masknet}        & 0.8125 & 0.4397  & 0.9679 & 0.2425  & 0.8142 & 0.5153  & 0.9832 & 0.1916  \\
2022 & GraphFM \cite{2021GraphFM}         & 0.8113 & 0.4405  & 0.9595 & 0.2384  & 0.8104 & 0.5189  & 0.9471 & 0.2665  \\
2023 & CL4CTR \cite{2023Cl4ctr}          & 0.8135 & 0.4383  & 0.9683 & 0.2148  &    0.8133    &  0.5180       & 0.9827 & 0.1559  \\ \hline
\multirow{2}{*}{ours} & \textbf{FSDNet}                 & \textbf{0.8149}       & \textbf{0.4371 }      &\textbf{ 0.9733}       & \textbf{0.1917}       & \textbf{0.8190}       &\textbf{ 0.5087 }     & \textbf{0.9856}       & \textbf{0.1367}      \\ \cline{2-10} 
     & T-test ($p$-values) & 2.37E-6      & 3.59E-6       & 1.17E-7      & 7.89E-9       & 2.02E-6      & 3.27E-7       & 1.84E-4      & 2.62E-3       \\ \hline
     \Xhline{1px}
\end{tabular}}
\label{baseline}
\end{table*}

\item The performance gap between baseline models is larger on smaller datasets like ML-1M and Frappe, whereas the gap is smaller on larger datasets such as Criteo. This is because larger datasets typically contain richer samples and feature information, and the increase in training data helps models better capture patterns within the data. When the dataset size is sufficiently large, all baseline models can learn from the ample data, and the model differences are smoothed out by the redundancy in the large-scale data. Therefore, on large datasets, the performance gap between various models becomes smaller. In contrast, on smaller datasets, the limited number of available training samples restricts the generalization ability of models, and factors like model complexity and different regularization strategies will significantly affect model performance. Consequently, model parameter settings need to be adjusted for different datasets to achieve optimal performance.

\item  FSDNet consistently outperforms the baseline models across all four datasets.
It makes significant progress compared to the strongest baseline models, with improvements in AUC of 0.11\%, 0.32\%, 0.55\%, and 0.11\%, and in Logloss of 0.16\%, 7.70\%, 1.20\%, and 5.98\% on the Criteo, ML-tag, ML-1M, and Frappe datasets, respectively. All these improvements exceed the significance threshold of 
\textit{\textbf{0.1\%}}  for the CTR direction, demonstrating the effectiveness of our proposed framework in CTR prediction tasks.  Ten experiments with random seeds are carried out, followed by a two-tailed T-test ($p$-value) to compare FSDNet with the best baseline model.
 The resulting $p$-values, being less than
$0.01$ \cite{2023GDCN}, demonstrate that the model's improvements are \textbf{statistically significant}.

\item Compared to the original DCNv2, FSDNet achieves improvements in AUC by 0.12\%, 0.33\%, 0.55\%, and 0.11\%, and in Logloss by 0.32\%, 14.95\%, 2.17\%, and 13.54\%, respectively. We attribute these significant improvements to two primary factors: (1) By introducing self-distillation, the soft labels from the teacher model are used to progressively guide the student model's optimization. With direct feedback, the student model can fine-tune its learning process more efficiently, capturing subtle nuances that might otherwise be missed, thereby enhancing both the model's generalization ability and overall performance.
(2) We separately  concatenate the layers of the cross  and the deep network, optimizing them through backpropagation.  This effectively resolves the issue of insufficient explicit and implicit feature interaction in parallel structure models.

\end{itemize}

\subsection{Compatibility Analysis (RQ2)}
We propose a lightweight and model-agnostic fusion self-distillation module, designed to be seamlessly integrated as a plug-and-play component into various parallel structure CTR models to enhance their performance. Here, we  examine  the compatibility of this module across multiple  models.

\begin{table}[t]
\centering
\renewcommand\arraystretch{1.3}
\caption{Study on the compatibility of the fusion self-distillation module. AUC-Imp and Logloss-Imp respectively represent the average improvement in AUC performance and Logloss performance of the model across four datasets. Typically, CTR researchers consider an increase of \textit{\textbf{0.1\%}} to be statistically significant.} 
\label{Compatibility study}
\resizebox{0.9\textwidth}{!}{ 
\begin{tabular}{c|cc|cc|cc|cc|c|c}
\hline
\Xhline{1px}
\multirow{2}{*}{Model} & \multicolumn{2}{c|}{Criteo} & \multicolumn{2}{c|}{ML-tag} & \multicolumn{2}{c|}{ML-1M} & \multicolumn{2}{c|}{Frappe} & \multirow{2}{*}{AUC-Imp} & \multirow{2}{*}{Logloss-Imp} \\ \cline{2-9}
                       & AUC          & Logloss      & AUC          & Logloss      & AUC         & Logloss      & AUC          & Logloss      &                              &                                  \\ \hline
DCN                    & 0.8138       & 0.4383       & 0.9702       & 0.2238       & 0.8143      & 0.5162       & 0.9838       & 0.1544       & \multirow{2}{*}{0.23\%}      & \multirow{2}{*}{5.90\%}          \\
DCN$_{FSD}$       & 0.8141       & 0.4379       & 0.9729       & 0.2006       & 0.8182      & 0.5109       & 0.9852       & 0.1357       &                              &                                  \\ \hline
xDeepFM                & 0.8140       & 0.4382       & 0.9692       & 0.2110       & 0.8142      & 0.5149       & 0.9844       & 0.1454       & \multirow{2}{*}{0.27\%}      & \multirow{2}{*}{3.08\%}          \\
xDeepFM$_{FSD}$    & 0.8142       & 0.4381       & 0.9728       & 0.1991       & 0.8187      & 0.5103       & 0.9855       & 0.1370       &                              &                                  \\ \hline
AutoInt+               & 0.8140       & 0.4378       & 0.9698       & 0.2266       & 0.8138      & 0.5182       & 0.9841       & 0.1520       & \multirow{2}{*}{0.26\%}      & \multirow{2}{*}{6.00\%}          \\
AutoInt+$_{FSD}$             & 0.8143       & 0.4377       & 0.9727       & 0.2030       & 0.8186      & 0.5099       & 0.9853       & 0.1338       &                              &                                  \\ \hline
DCNv2                 & 0.8139       & 0.4385       & 0.9701       & 0.2254       & 0.8145      & 0.5200       & 0.9845       & 0.1581       & \multirow{2}{*}{0.28\%}      & \multirow{2}{*}{7.74\%}          \\
DCNv2$_{FSD}$            & 0.8149       & 0.4371       & 0.9733       & 0.1917       & 0.8190      & 0.5087       & 0.9856       & 0.1367       &                              &                                  \\ 
\hline
\Xhline{1px}
\end{tabular}}
\end{table}

\begin{table}[t]
\small
\renewcommand\arraystretch{1.1}
\caption{Parameter analysis of different models before and after adding the fusion self-distillation module.} 
\label{Parameter analysis}
\centering
\resizebox{0.55\linewidth}{!}{
\begin{tabular}{c|cccc}
\Xhline{1px}
\hline
Model      & Criteo   & ML-tag  & ML-1M  & Frappe \\ \hline
DCN        & 15,149,921 & 1,761,073 & 510,577 & 475,249 \\
DCN$_{FSD}$     & 15,152,996 & 1,762,420 & 512,020 & 476,932 \\ \hline
xDeepFM    & 16,473,517 & 1,874,870 & 562,588 & 568,217 \\
xDeepFM$_{FSD}$ & 16,474,511 & 1,875,864 & 563,582 & 569,211 \\ \hline
AutoInt+   & 16,087,469 & 1,956,214 & 626,780 & 587,289 \\
AutoInt+$_{FSD}$ & 16,095,759 & 1,958,168 & 629,502 & 591,931 \\ \hline
DCN-v2     & 16,316,177 & 1,767,841 & 529,537 & 551,569 \\
DCNv2$_{FSD}$   & 16,319,252 & 1,769,188 & 530,980 & 553,252 \\ \hline
\Xhline{1px}
\end{tabular}}
\vspace{-1em}
\end{table}

 We conduct extensive experiments by integrating the fusion self-distillation module into four classic parallel structure CTR models: xDeepFM \cite{2018xdeepfm}, AutoInt \cite{2019autoint}+, DCN \cite{2017DCN}, and DCNv2 \cite{2021dcnv2}. For xDeepFM, we concatenate the high-order interactions constructed by each layer of the CIN (Compressed Interaction Network) with the  output of corresponding deep network layer. These features are then used together for subsequent prediction tasks and optimized through self-distillation. For AutoInt+, the output of each layer from the multi-head self-attention module is concatenated with the output of the corresponding layer in the deep network. The concatenated tensor includes both the feature relevancies extracted by the self-attention mechanism and the implicit feature interactions derived from the deep network layers. For DCN and DCNv2,  we concatenate the outputs of each layer in the cross network with the corresponding layer in the deep network.  To facilitate distinction, we denote models equipped with the fusion  self-distillation module as $\mathcal{M}_{FSD}$. It should be noted that DCNv2$_{FSD}$ is actually the FSDNet proposed by us.

As shown in Table \ref{Parameter analysis}, the increase in model parameters is relatively small after introducing the fusion self-distillation module. Therefore, this module can be considered a lightweight enhancement approach.
From the Table \ref{Compatibility study}, it is evident that the integration of this module enhances the performance of the models. For DCN, xDeepFM, AutoInt+ and DCNv2 across four datasets, the average improvements relative to the AUC metric are 0.23\%, 0.27\%, 0.26\%, 0.28\%, and for the Logloss metric, they are 5.90\%, 3.08\%, 6.00\%, 7.74\%.  In the field of CTR, an improvement of \textit{\textbf{0.1\%}} is considered significant, which demonstrates the effectiveness of feature information fusion and self-distillation during model optimization. The underlying reason is that feature information fusion addresses the issue of insufficient utilization of feature sets due to lack of communication between different paths, which is often overlooked in parallel structure models. Furthermore, the introduction of self-distillation allows the student model to receive guidance from the teacher model at each layer, facilitating more effective adjustments. The significant improvement in Logloss demonstrates the particularly notable effect of the self-distillation module in enhancing classification tasks.

\subsection{Hyperparameter Study (RQ3)}
In this part, we further examine the influence of several key hyperparameters on the FSDNet model, including the loss balance parameter $\mu$, the temperature $\tau$, and the hint loss weight $\gamma$.
\subsubsection{Impact of Loss Balance $\mu$}
\begin{figure*}[t]
\vspace{-1em}
    \subfloat{
        \centering
        \includegraphics[width=0.33\linewidth]{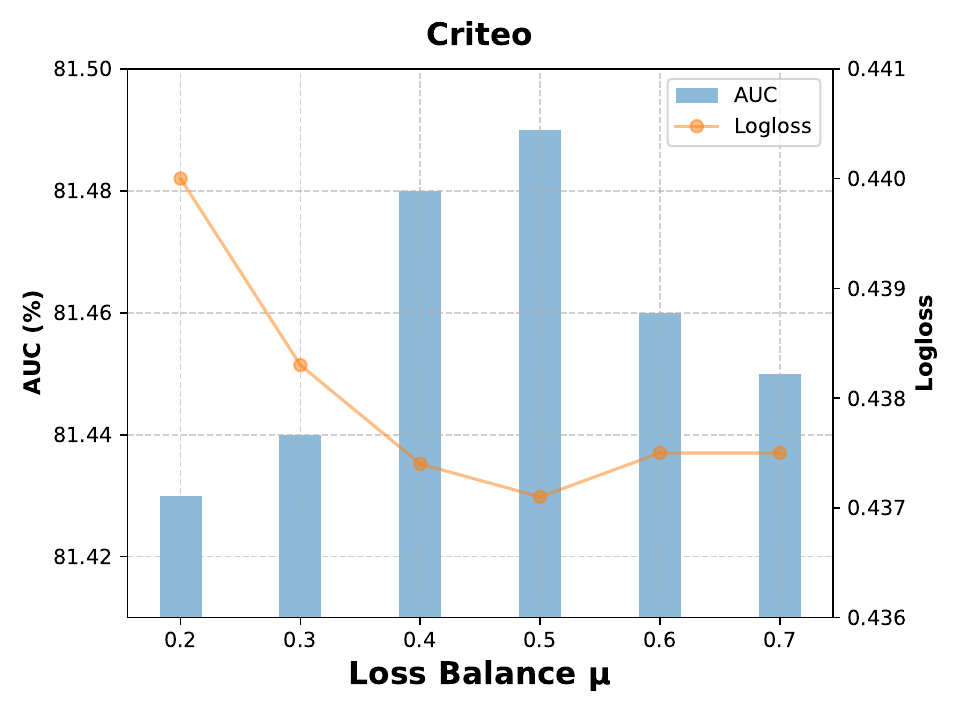}%
    }\hspace{-2.5mm}
    \subfloat{
        \centering
        \includegraphics[width=0.33\linewidth]{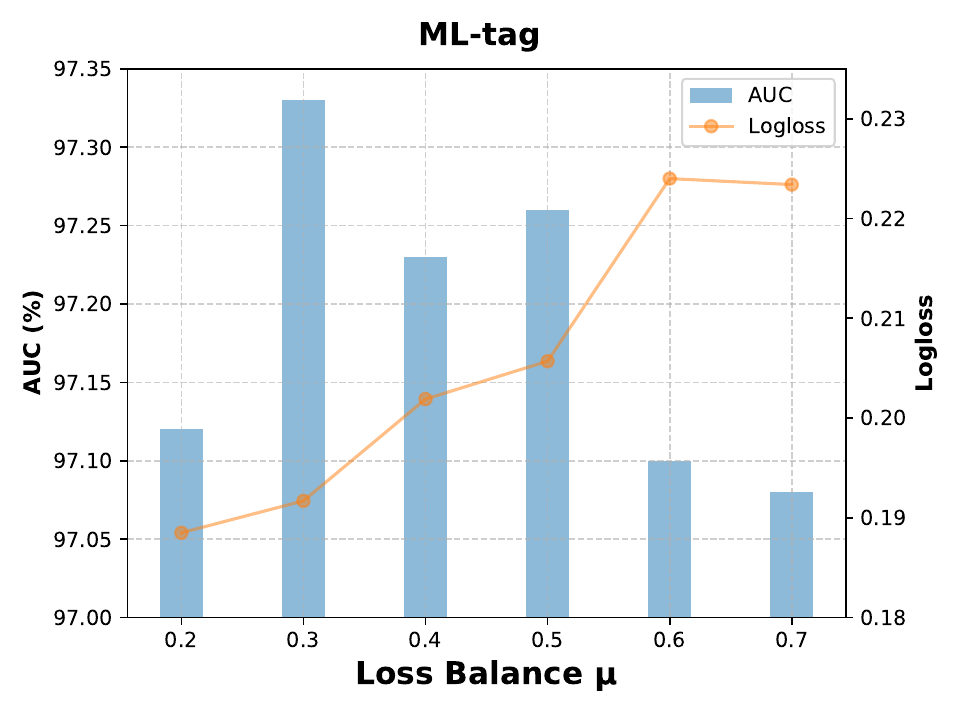}
    }\hspace{-3.1mm}
    \subfloat{
        \centering
        \includegraphics[width=0.33\linewidth]{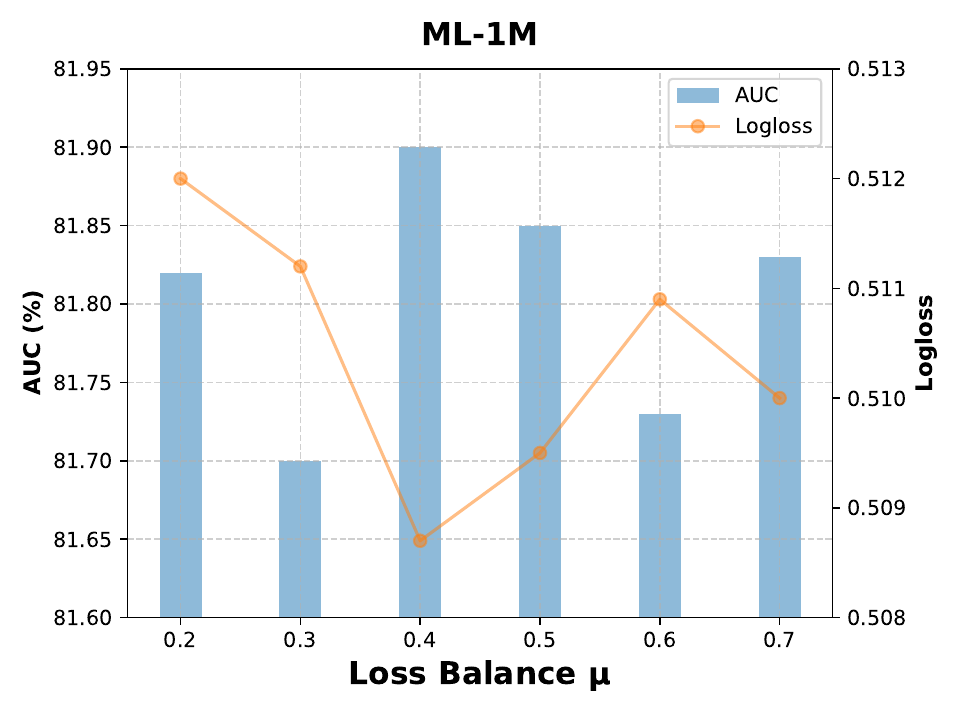}
    }
    \captionsetup{justification=raggedright}
    \caption{Hyperparameter study of Loss Balance $\mu$.}
    \label{Loss Balance}
\end{figure*}

We fix all other hyperparameters and adjust $\mu$ with a step size of 0.1.  As indicated in Figure \ref{Loss Balance}, the model reaches optimal performance at $\mu=0.5$ on Criteo, $\mu=0.3$ on ML-tag, and $\mu=0.4$ on ML-1M. The optimal values of $\mu$ vary across different datasets, indicating that the loss balance setting should be adjusted  based on the specific application context and data characteristics. Generally, the model performs optimally  with the $\mu$ value around $0.5$, demonstrating that during training, hints from soft labels and true labels are equally important.

\begin{figure*}[t]
\vspace{-1em}
    \subfloat{
        \centering
        \includegraphics[width=0.33\linewidth]{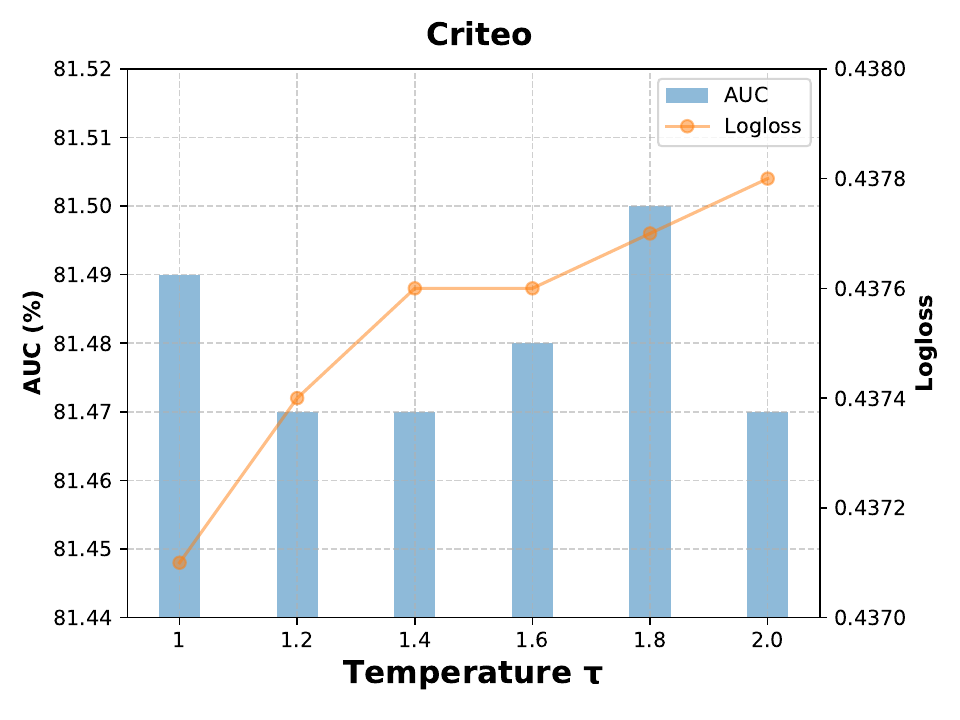}%
    }\hspace{-2.5mm}
    \subfloat{
        \centering
        \includegraphics[width=0.33\linewidth]{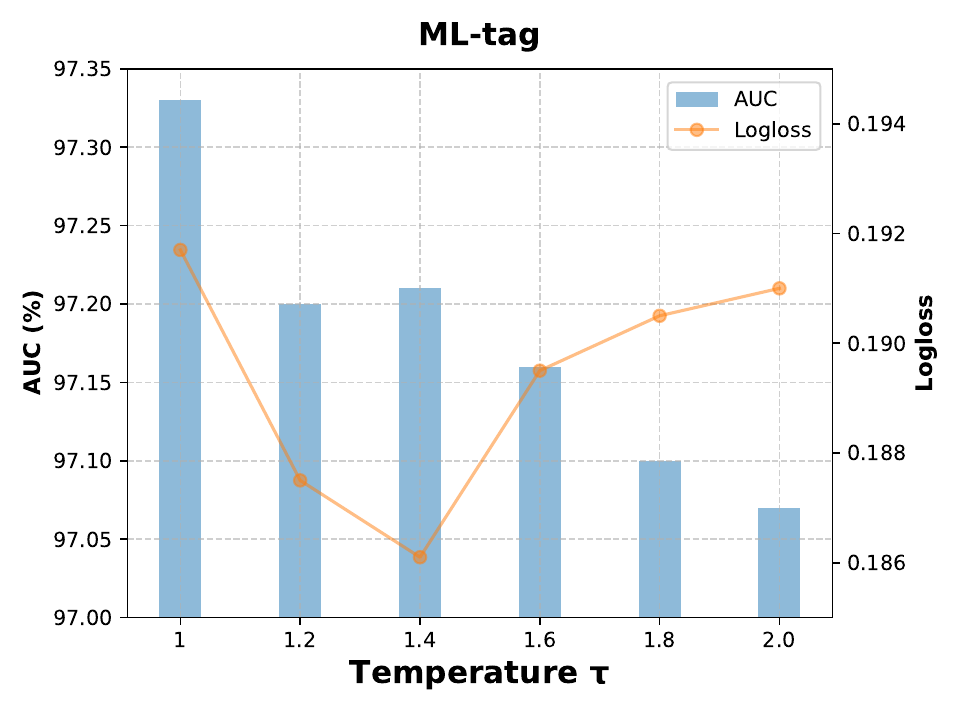}
    }\hspace{-3.1mm}
    \subfloat{
        \centering
        \includegraphics[width=0.33\linewidth]{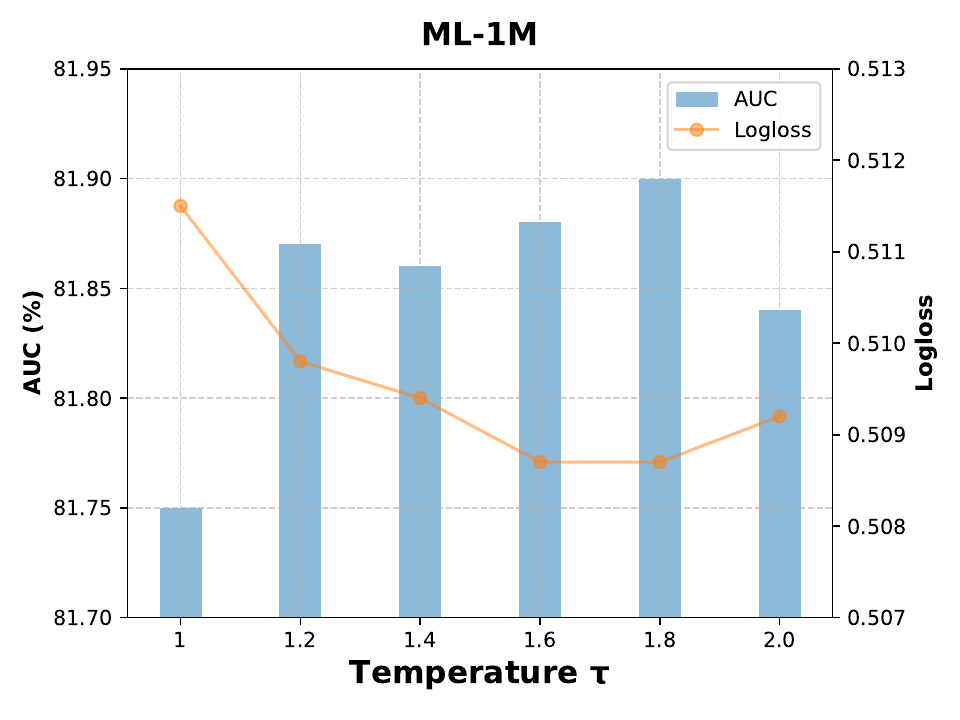}
    }
    \captionsetup{justification=raggedright}
    \caption{Hyperparameter study of  Temperature $\tau$.}
    \label{Temperature}
    \vspace{-0.5em}
\end{figure*}

\begin{figure*}[t]
\vspace{-0.5em}
    \subfloat{
        \centering
        \includegraphics[width=0.33\linewidth]{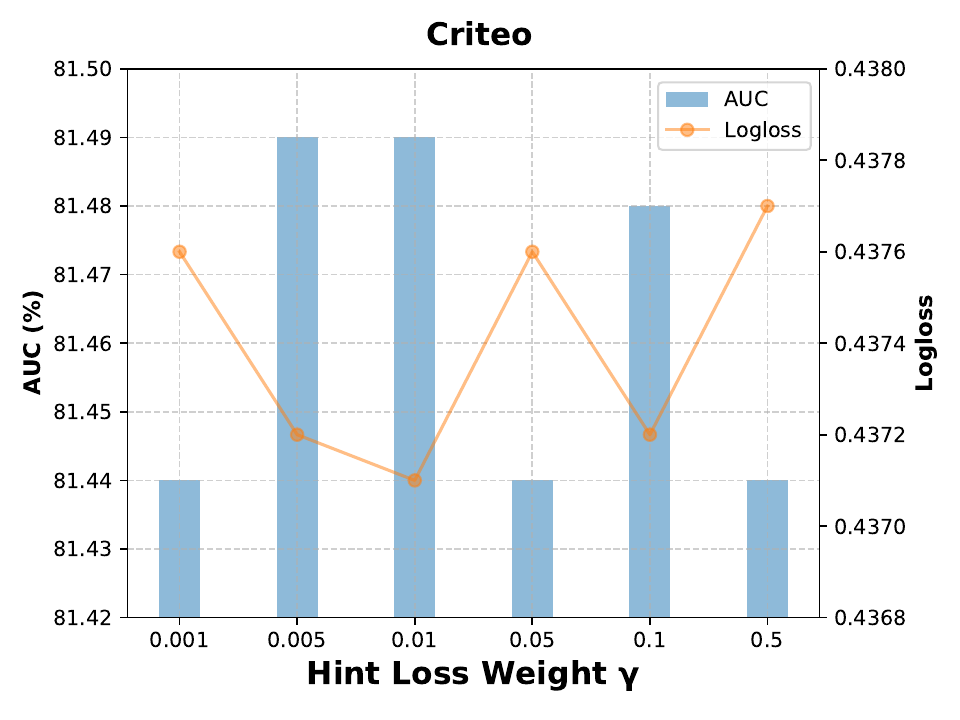}%
    }\hspace{-2.5mm}
    \subfloat{
        \centering
        \includegraphics[width=0.33\linewidth]{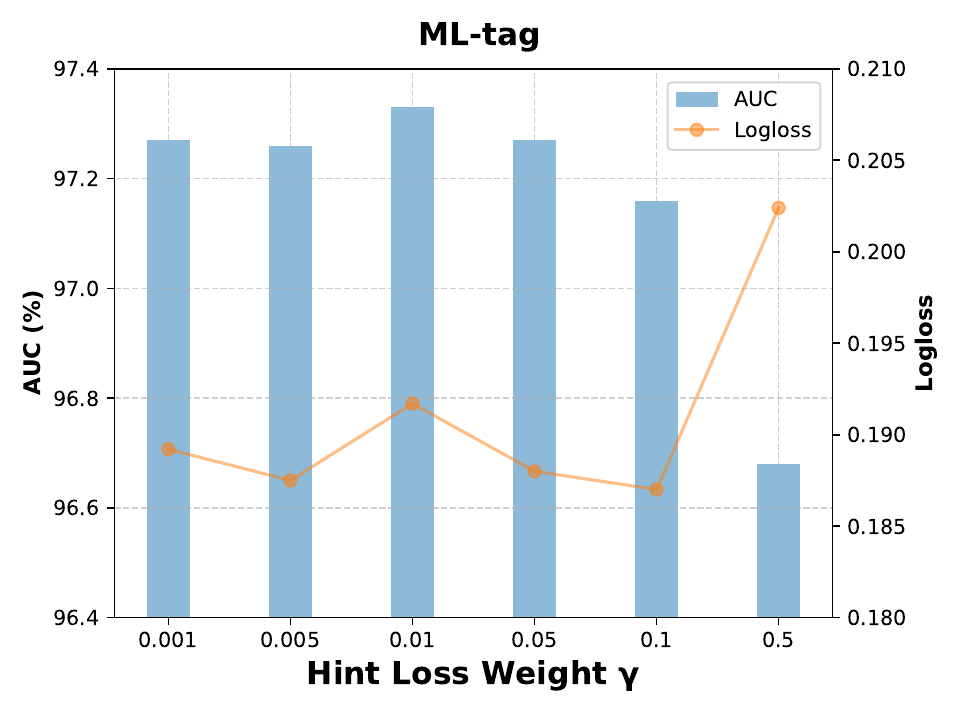}
    }\hspace{-3.1mm}
    \subfloat{
        \centering
        \includegraphics[width=0.33\linewidth]{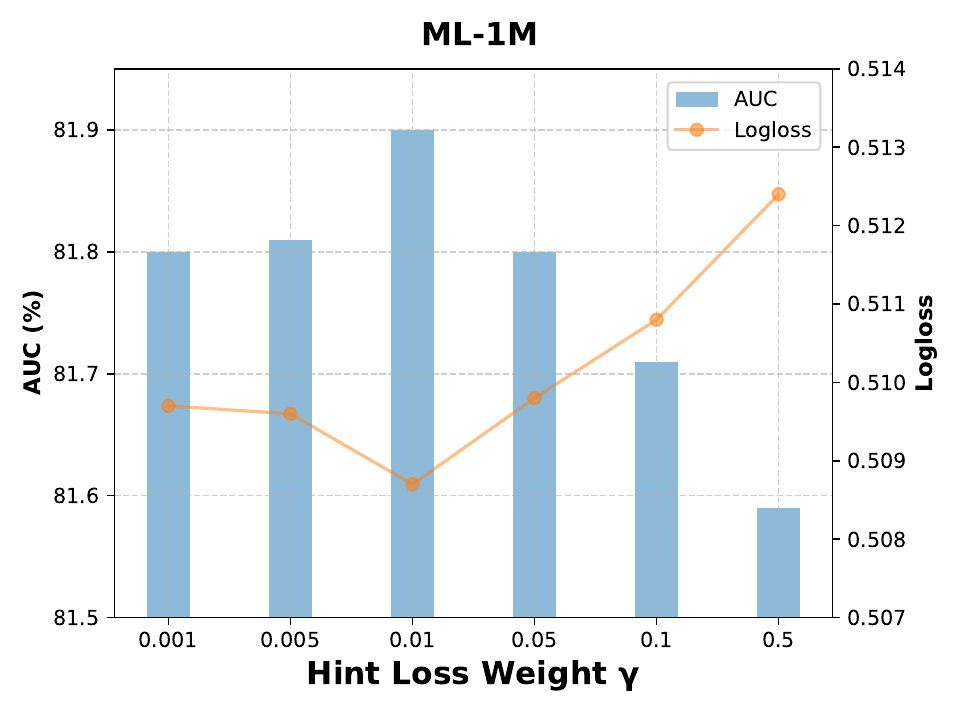}
    }
    \captionsetup{justification=raggedright}
    \caption{Hyperparameter study of  Hint Loss Weight $\gamma$.}
    \label{Hint Loss}
\end{figure*}

\subsubsection{Impact of Temperature $\tau$}
Different datasets have varying sensitivities to temperature. As shown in Figure \ref{Temperature}, the AUC for the Criteo is optimal at $\tau = 1.8$, while the lowest Logloss occurs at $\tau = 1$. For ML-tag and ML-1M, the AUC performs best at $\tau = 1.0$ and $\tau = 1.8$, respectively. Generally, increasing the temperature can reduce model overfitting, making the model smoother. However too high a temperature might also lead to a decline in the model's discriminative ability, as indicated by a decrease in Logloss performance.
\subsubsection{Impact of Hint Loss Weight $\gamma$}
As shown in Figure \ref{Hint Loss}, the Criteo, ML-tag, and ML-1M all achieve optimal performance when $\gamma$ is set to 0.01. Both excessively high (e.g., 0.5) and excessively low (e.g., 0.001) values are inappropriate. A high hint loss weight can interfere with the learning of the main task, leading to overfitting. On the other hand, too low a value may fail to adequately guide the student model to learn the teacher model's effective features. This would fail to achieve the primary advantage of self-distillation, which is leveraging the teacher model's internal knowledge to improve  the student model's performance.

\subsection{Ablation Study (RQ4)}
We perform  ablation studies to explore the impact  of each design module within the model, considering the following three variants:
\begin{itemize}[leftmargin=*]
\item \textbf{w/o SL}: We removed the supervision of soft labels in the self-distillation process.
\item \textbf{w/o HL}: We do not implement hint loss during the self-distillation process.
\item \textbf{w/o IF}: We eliminated the part that constructs explicit and implicit feature interaction information fusion, focusing solely on self-distillation of the deep network.
\end{itemize}

\begin{table}[t]
\small
\renewcommand\arraystretch{1}
\centering
\caption{Ablation study of FSDNet.} 
\label{Ablation study}
\resizebox{0.85\linewidth}{!}{
\begin{tabular}{c|cc|cc|cc|cc}
\Xhline{1px}
\hline
\multirow{2}{*}{Model} & \multicolumn{2}{c|}{Criteo} & \multicolumn{2}{c|}{ML-tag} & \multicolumn{2}{c|}{ML-1M} & \multicolumn{2}{c}{Frappe} \\ \cline{2-9} 
       & AUC    & Logloss & AUC    & Logloss & AUC    & Logloss & AUC    & Logloss \\ \hline
w/o SL & 0.8146 & 0.4376  & 0.9711 & 0.2464  & 0.8181 & 0.5111  & 0.9841 & 0.1409  \\
w/o HL & 0.8146 & 0.4375  & 0.9723 & 0.1899  & 0.8184 & 0.5091  & 0.9843 & 0.1411  \\
w/o IF & 0.8147 & 0.4375  & 0.9720 & 0.2015  & 0.8166 & 0.5159  & 0.9845 & 0.1546  \\ \hline
FSDNet & 0.8149 & 0.4371  & 0.9733 & 0.1917  & 0.8190 & 0.5087  & 0.9856 & 0.1367  \\ \hline
\Xhline{1px}
\end{tabular}}
\end{table}

As shown in Table \ref{Ablation study}, both the {w/o SL model and the w/o HL model exhibit a decline in performance. It is evident that although both loss functions play an important role in improving model performance, the Soft Labels loss contributes a bit more than the Hint Loss. The Soft Labels loss boosts the generalization ability of the student model by forcing it to learn the output distribution of the teacher model, whereas the Hint Loss improves the feature learning capacity of the student model by explicitly conveying intermediate feature layers. The findings from the ablation experiments confirm the effectiveness of these two losses in improving model performance, while also revealing their complementary roles within the self-distillation framework. 

Furthermore, the disparity in performance between the w/o IF model and the FSDNet model indicates that removing the information fusion part significantly  degrades the FSDNet model’s performance. This highlights the issue that independently constructing explicit and implicit feature interactions leads to a suboptimal utilization of feature information.

\subsection{Deep and Cross Layer Combination Methods (RQ5)}
\label{Combination}
Here, we seek methods to associate  each layer of the cross network with the corresponding layer of the deep network, defined by the function $\mathbf{f}_l=f(\mathbf{c}_l, \mathbf{d}_l)$, where $\mathbf{c}_l$ and $\mathbf{d}_l$ denote the outputs of the $l^{th}$ layer of the cross network and deep network, correspondingly, and $f(\cdot)$  is a predefined interaction function. Based on empirical studies, we compared four methods of interaction.

\begin{figure}[t]
    \centering
    \begin{minipage}[t]{0.25\linewidth}
        \centering
        \includegraphics[width=\textwidth]{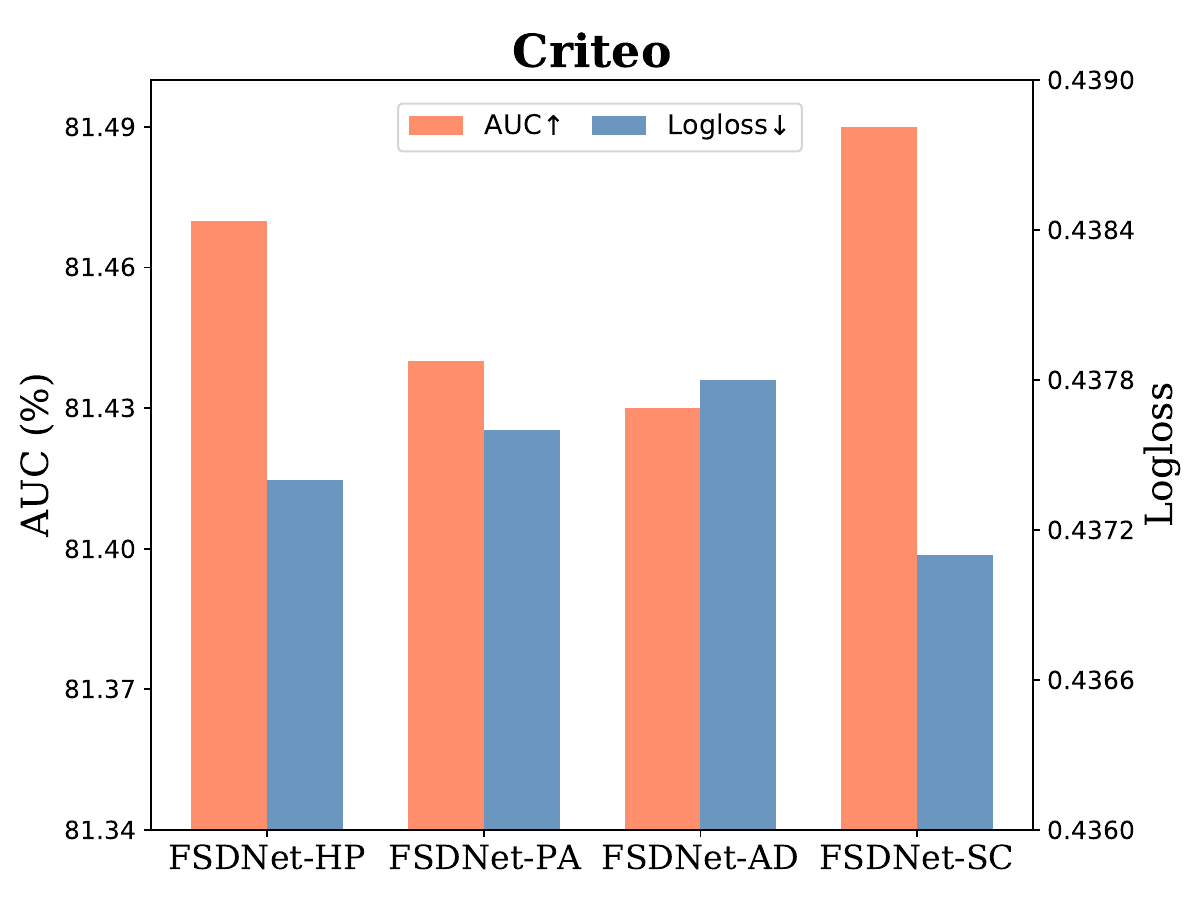}
        \centerline{(a) Criteo}
    \end{minipage}%
    \begin{minipage}[t]{0.25\linewidth}
        \centering
        \includegraphics[width=\textwidth]{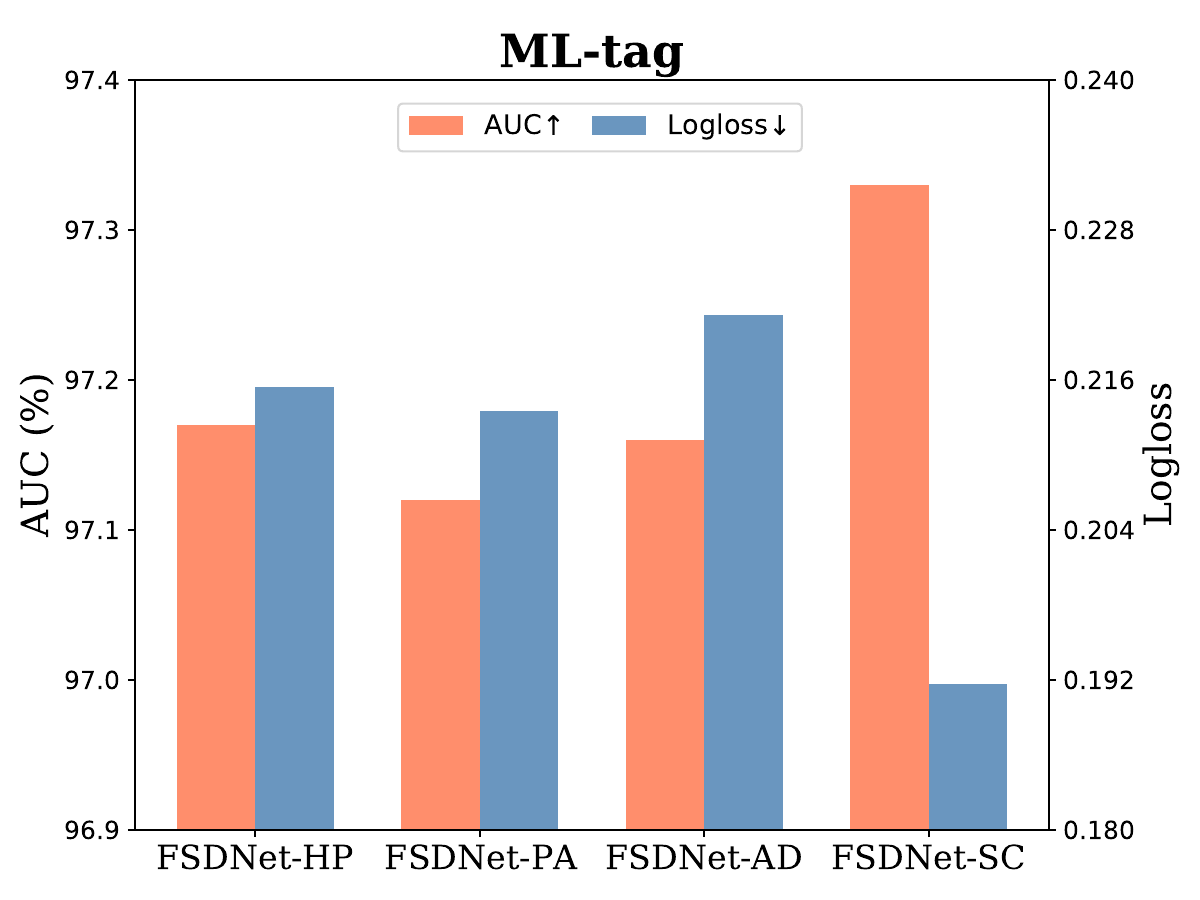}
        \centerline{(b) ML-tag}
    \end{minipage}%
    \begin{minipage}[t]{0.25\linewidth}
        \centering
        \includegraphics[width=\textwidth]{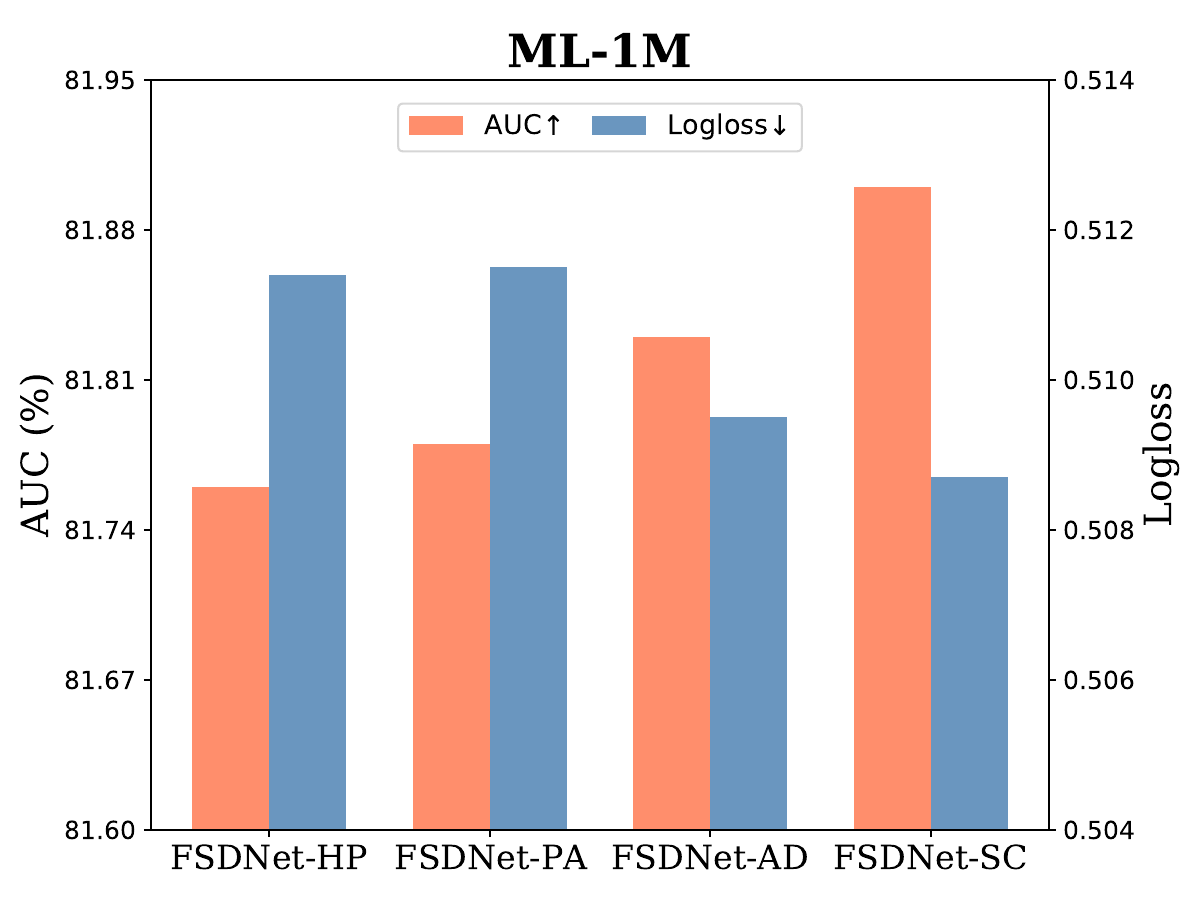}
        \centerline{(c) ML-1M}
    \end{minipage}%
    \begin{minipage}[t]{0.25\linewidth}
        \centering
        \includegraphics[width=\textwidth]{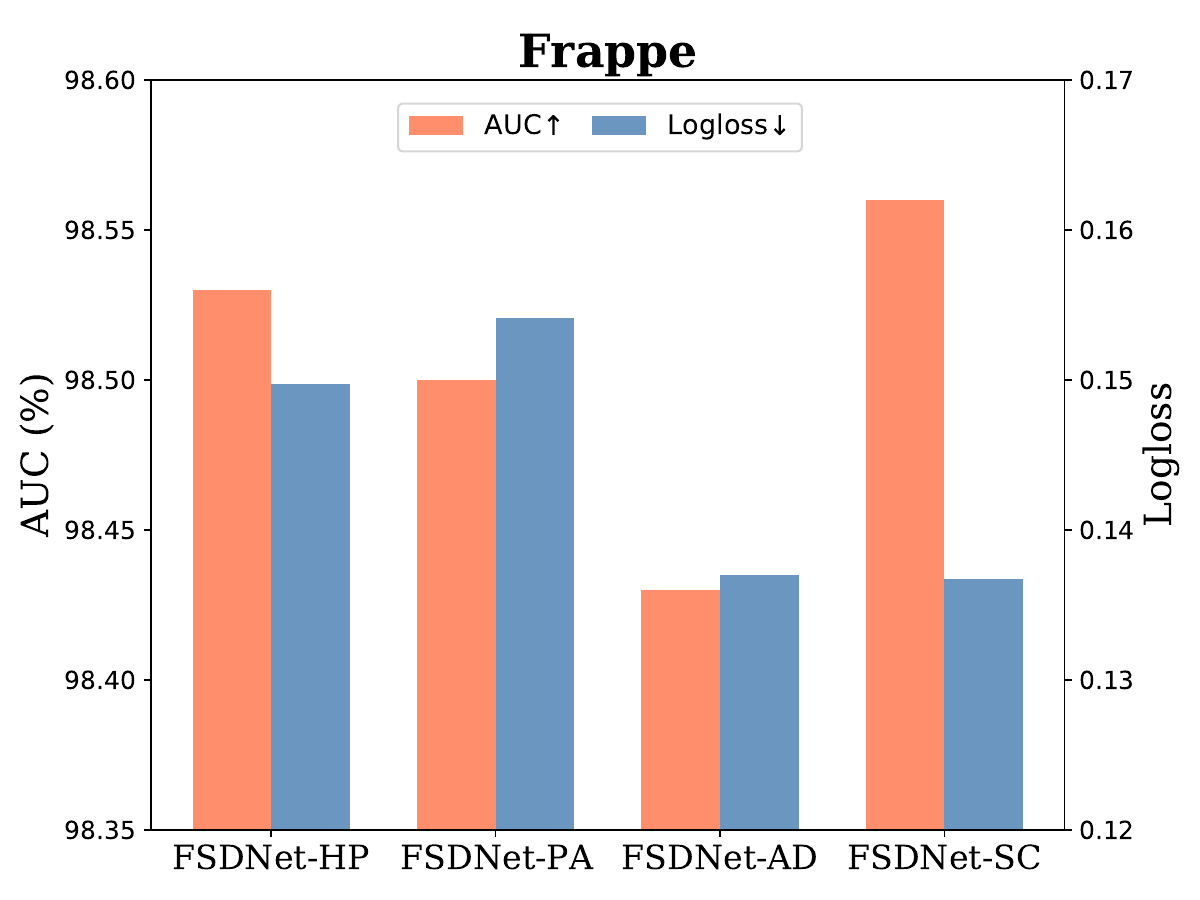}
        \centerline{(d) Frappe}
    \end{minipage}%
    \captionsetup{justification=raggedright}
    \caption{Study on deep and cross layer combination methods.}
    \label{connection}
    \vspace{-1em}
\end{figure}

     \textit{Simple Concatenation (SC):} Directly concatenate the outputs of each layer.  No additional parameters are introduced, and the concatenation layer is represented as: $\mathbf{f}_l=[\mathbf{c}_l; \mathbf{d}_l]$. 
     \textit{Hadamard Product (HP):} The $\mathbf{d}_l$ is transformed via a linear layer into a result $\mathbf{d}_t$ with the same dimension as $\mathbf{c}_l$,  followed by a Hadamard Product (i.e., element-wise multiplication), introducing deeper interactions.
    The result is then concatenated with $\mathbf{d}_t$ and $\mathbf{c}_l$. It is represented as:
    $\mathbf{f}_l= [\mathbf{c}_l \otimes \mathbf{d}_t, \mathbf{c}_l, \mathbf{d}_t]$.
     \textit{Pointwise Addition(PA):} The result
    $\mathbf{d}_t$ is added element-wise with $\mathbf{c}_l$, followed by concatenation. It is represented as:    $\mathbf{f}_l= [\mathbf{c}_l \oplus \mathbf{d}_t, \mathbf{c}_l, \mathbf{d}_t]$.
     \textit{Attention Distribution (AD):} Utilizes the attention mechanism to capture the interaction relationships between input vectors \cite{2017attentionisneed,2021EDCN}. By adaptively assigning weights, the model can focus on more important feature interactions. And perform residual connection to avoid information loss. The corresponding formula is:  $\mathbf{f}_l=[\alpha_l\mathbf{c}_l+\mathbf{c}_l, \beta_l\mathbf{d}_l+\mathbf{d}_l $], where $\alpha_l$ and $\beta_l$ are attention weights in the $l^{th}$ layer. $\alpha_l = Softmax(\mathbf{p}_l^TReLU(\mathbf{W}_l^T\mathbf{c}_l+\mathbf{b_l)})$, where $\mathbf{p}_l^T$ is the trainable weight parameter used to calculate the attention scores, $\mathbf{W}_l^T$ and $\mathbf{b_l}$ are the weight matrix and bias vector, respectively. $\beta_l$ is computed in a similar manner.

 By changing only the connection methods, we establish different models:  simple concatenation (FSDNet-SC), hadamard product (FSDNet-HP), pointwise addition (FSDNet-PA), and attention distribution (FSDNet-AD). 
  The experimental results, as illustrated in Figure \ref{connection}, indicate that the model based on simple concatenation (FSDNet-SC) outperforms the others across all datasets. This superior performance can be attributed to FSDNet-SC's avoidance of additional parameters and operations, allowing it to retain the original feature information with minimal loss.
This enables the model to more effectively capture the complementary relationship  between feature interactions in the deep and cross networks during training.

In self-distillation, knowledge from the deepest network directly guides the shallower networks via backpropagation, maximizing input utilization and capturing both high-order and low-order interactions.  Introducing excessive complexity in the concatenation process may increase feature correlations,  but it also risks information loss or distortion, which can diminish the guidance effect. Additionally, these complex operations may lead to unstable gradient propagation, further impacting model stability.

\subsection{Visualization of each fusion layer (RQ6)}
To explore the impact of the fusion self-distillation module on feature representations at each layer, we compare FSDNet with DCNv2 using the Criteo dataset. We randomly sample the intermediate and final fusion layer representations for 4,096 instances. Then, we visualize these representations using t-SNE, as shown in Figure \ref{Visualization}.

\begin{figure*}[t]
    \subfloat[\footnotesize Intermediate layer (FSDNet)]{
        \centering
        \includegraphics[width=0.25\linewidth]{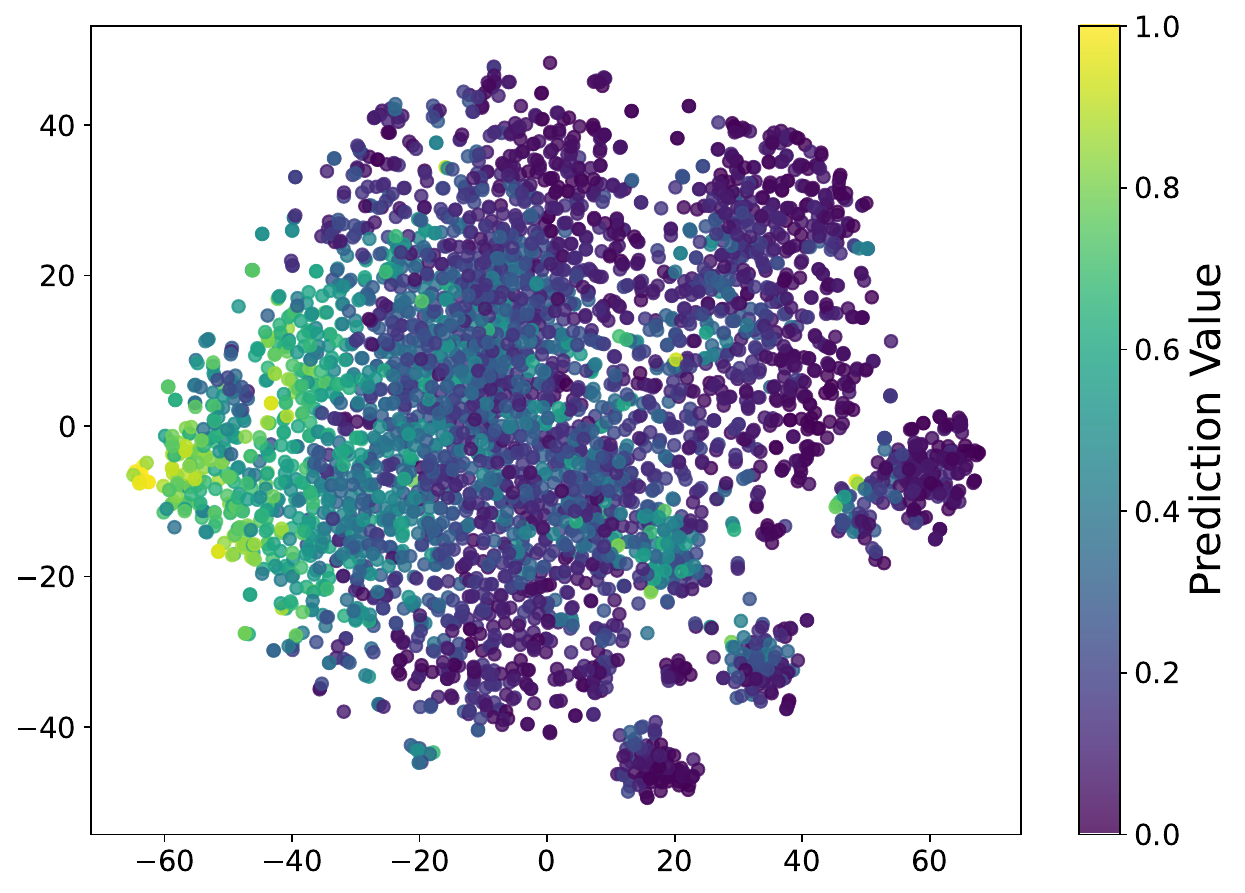}%
    }
    \subfloat[\footnotesize Final layer (FSDNet)]{
        \centering
        \includegraphics[width=0.25\linewidth]{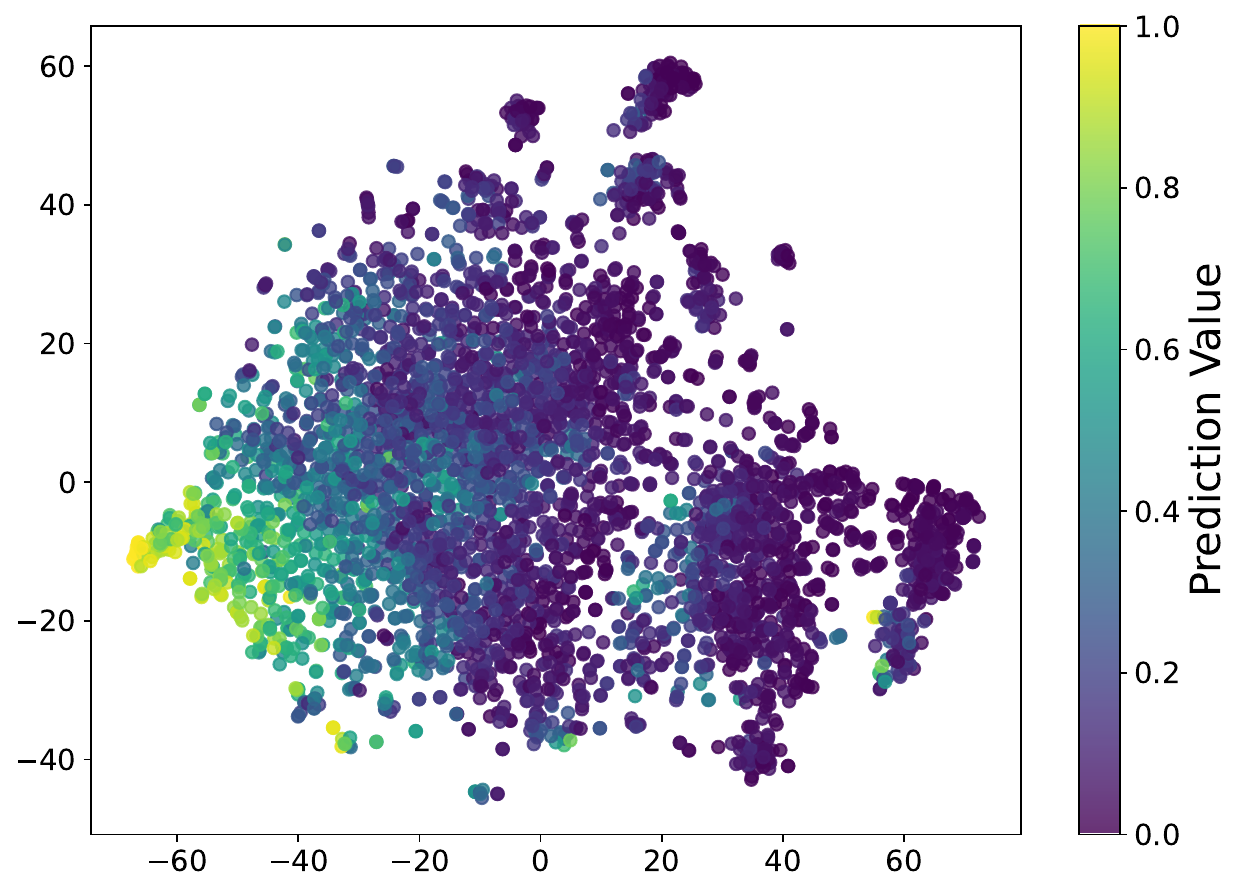}
    }
    \subfloat[\footnotesize Intermediate layer (DCNv2)]{
        \centering
        \includegraphics[width=0.25\linewidth]{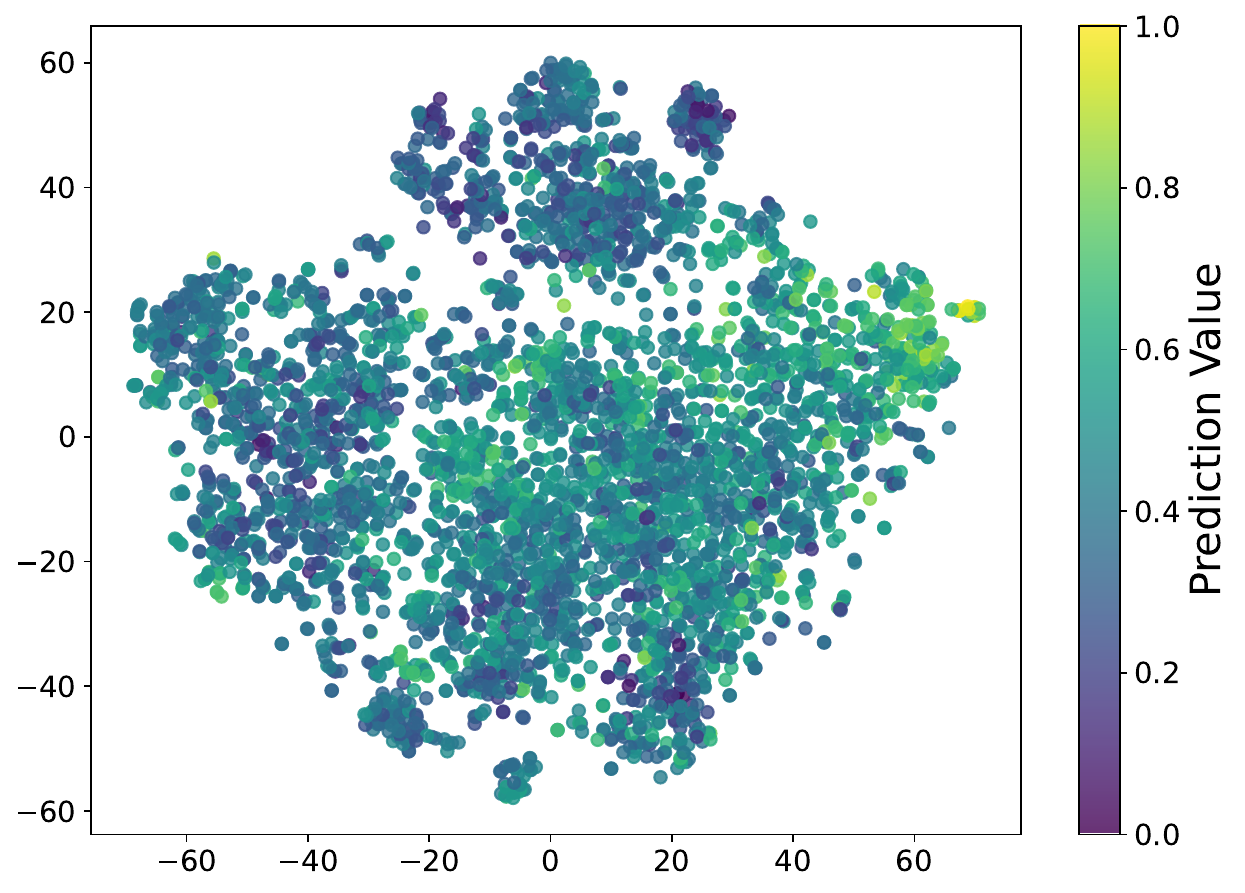}
    }
    \subfloat[\footnotesize Final layer (DCNv2)]{
        \centering
        \includegraphics[width=0.25\linewidth]{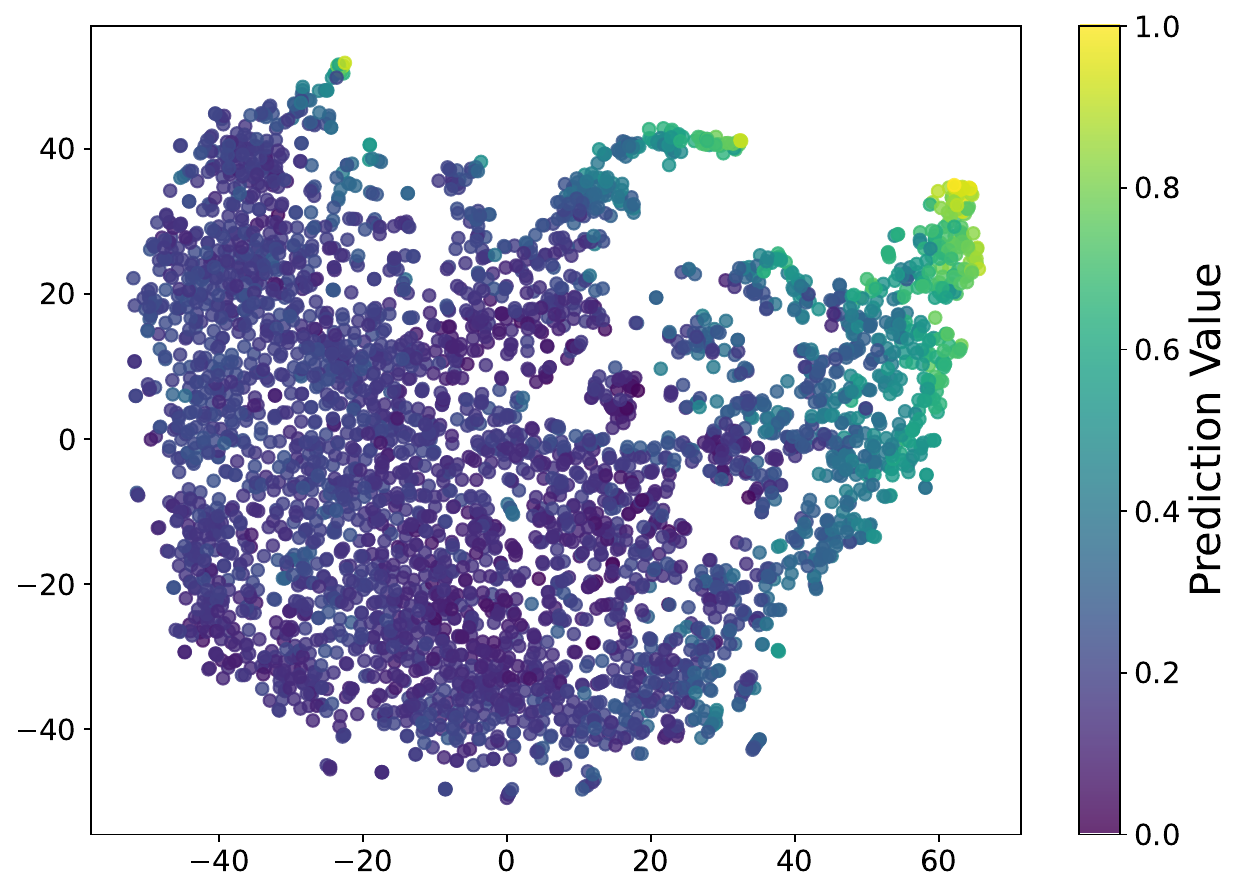}
    }
    
    \captionsetup{justification=raggedright}
    \caption{Visualization of each fusion layer representation on the Criteo dataset, with each point colored according to the prediction value.}
    \label{Visualization}
    \vspace{-1em}
\end{figure*}

It is clear that points with higher prediction values (brighter colors) tend to cluster in specific regions in the higher-level t-SNE plots. This indicates that the model effectively distinguishes high prediction value samples (positive samples as identified by the model).
FSDNet demonstrates a more pronounced ability for high prediction value clustering and feature differentiation. By using the deep layer output as a "teacher" to guide the learning of shallower layer features, it forms clearer feature clusters even at intermediate layers. This shows that the fusion self-distillation module progressively optimizes feature learning across different layers, enabling the model to capture useful feature interaction patterns earlier and achieve more precise feature extraction at the final layer.
In contrast, DCNv2 shows more dispersed feature distributions across layers, and the clustering effect of high prediction value points is less significant than in FSDNet. This reflects its limitations in feature interaction extraction and differentiation capabilities.

\subsection{Robustness to Noisy Interactions (RQ7)}
As mentioned above, in real-world CTR prediction, label data (such as whether a user clicks on an ad) often contains a lot of noise. Additionally, during the process of constructing feature interactions, numerous incorrect interactions may arise, further introducing noise. 
An effective model should not only perform well on clean data but also maintain stable performance in the presence of noisy data.  
FSDNet can effectively reduce the impact of noise introduction through the self-distillation mechanism.
To confirm the robustness of FSDNet against noisy interactions, we artificially introduced noise into the training set by modifying  specific proportions of the labels (i.e., $5\%$, $10\%$, $15\%$, $20\%$), while keeping the validation set and test set unchanged.  Figure \ref{Noise} shows the results on the Criteo, ML-tag, and ML-1M datasets.

\begin{itemize}
    \item We can clearly observe that as noise increases, the performance of the models is adversely affected. As the noise level increases from $0$ to $20\%$, the performance of all models declines. However, FSDNet consistently demonstrates the best performance, and the rate of decline for FSDNet is lower than that of DCN and DCNv2. This indicates that soft labels provide a smoother class distribution compared to hard labels. Even when some samples are mislabeled, the predictions from the teacher model can still offer more reliable guidance, effectively mitigating the impact of label noise. In short, FSDNet offers a different perspective for eliminating false interactions in the CTR prediction task.
    \item Focusing on the ML-1M dataset, FSDNet with $20\%$ noisy interactions still outperforms DCNv2 and DCN trained on noise-free datasets, further demonstrating the superiority and robustness of FSDNet.
    
\end{itemize}

\begin{figure}[t]
\vspace{-1em}
    \subfloat[Criteo]{
        \centering
        \includegraphics[width=0.33\linewidth]{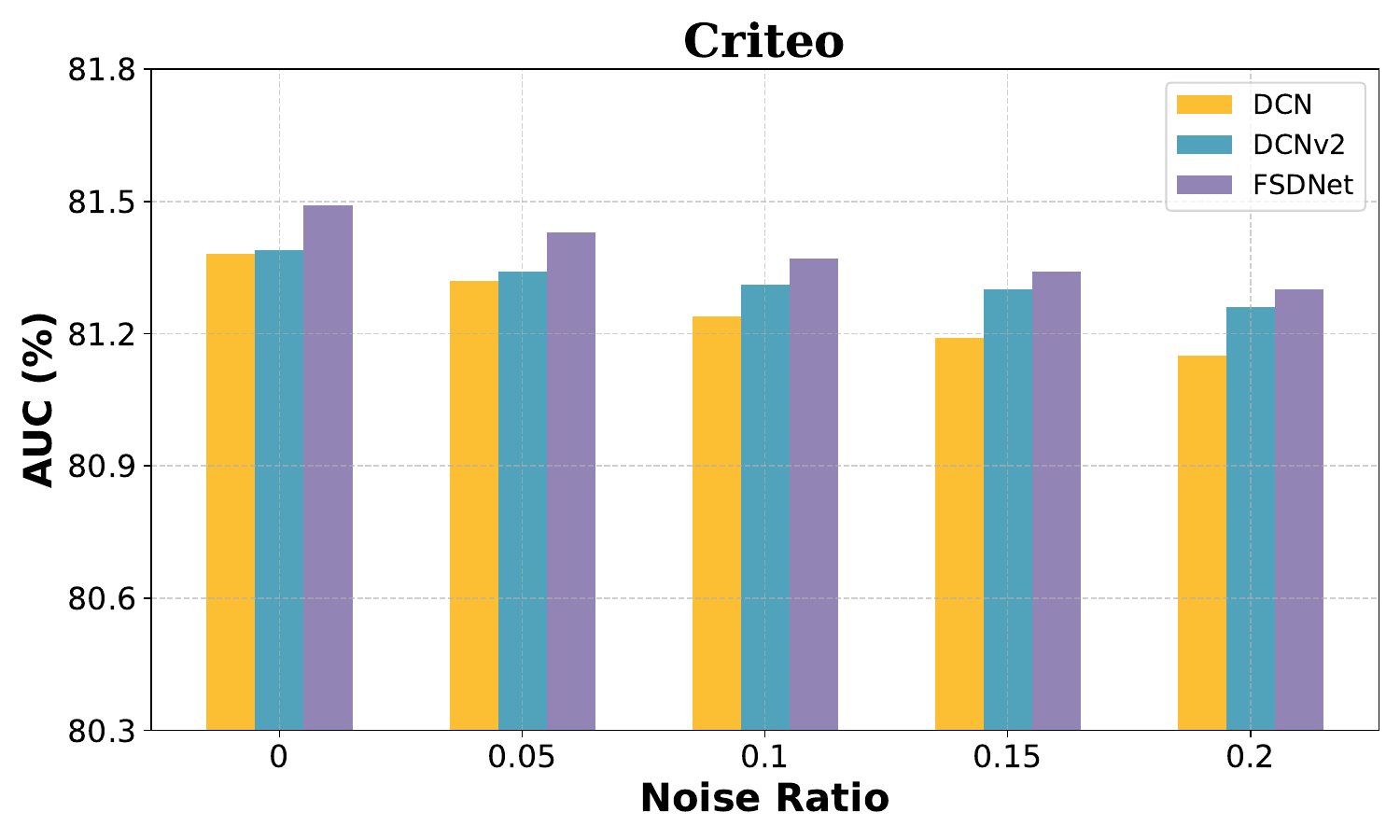}%
    }
    \subfloat[ML-tag]{
        \centering
        \includegraphics[width=0.33\linewidth]{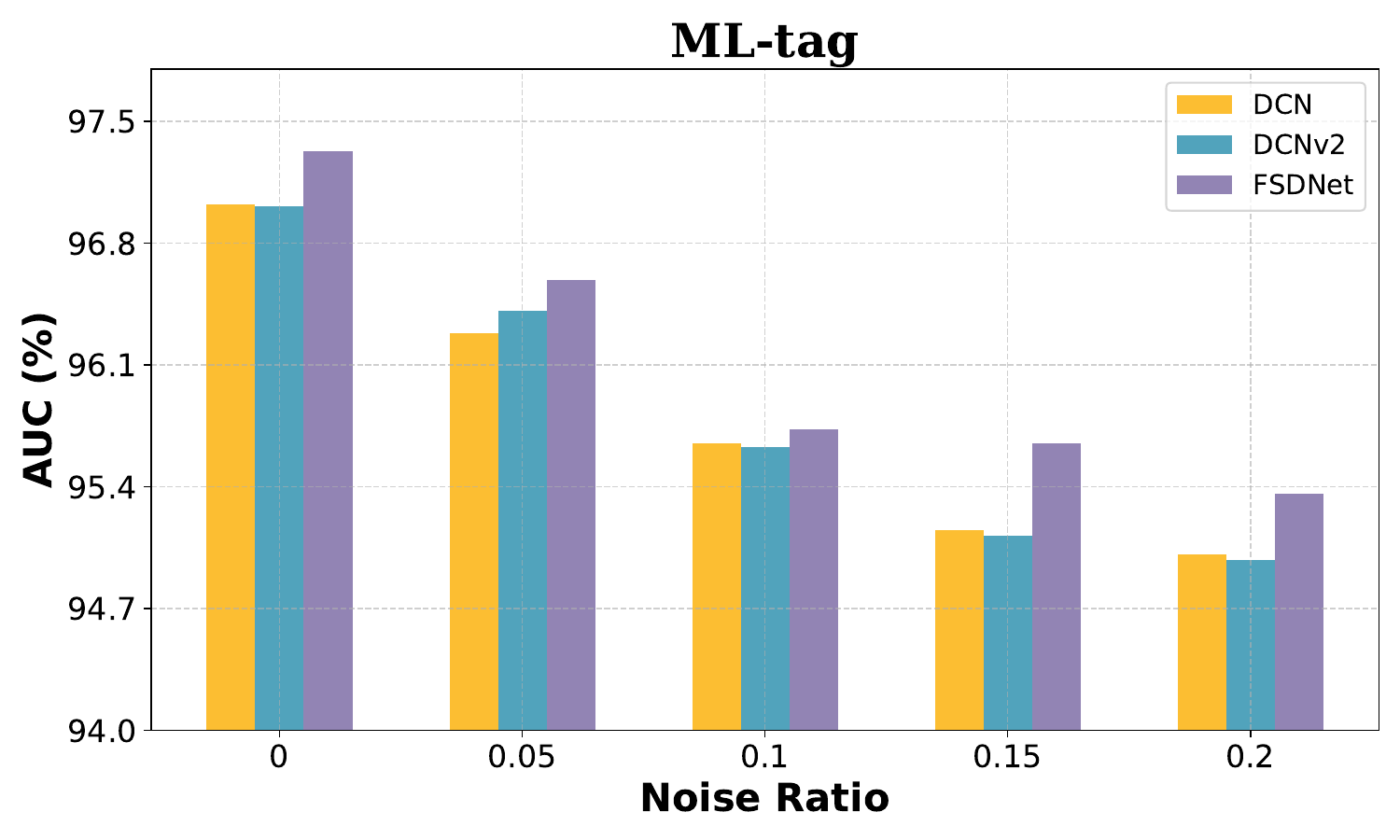}
    }
    \subfloat[ML-1M]{
        \centering
        \includegraphics[width=0.33\linewidth]{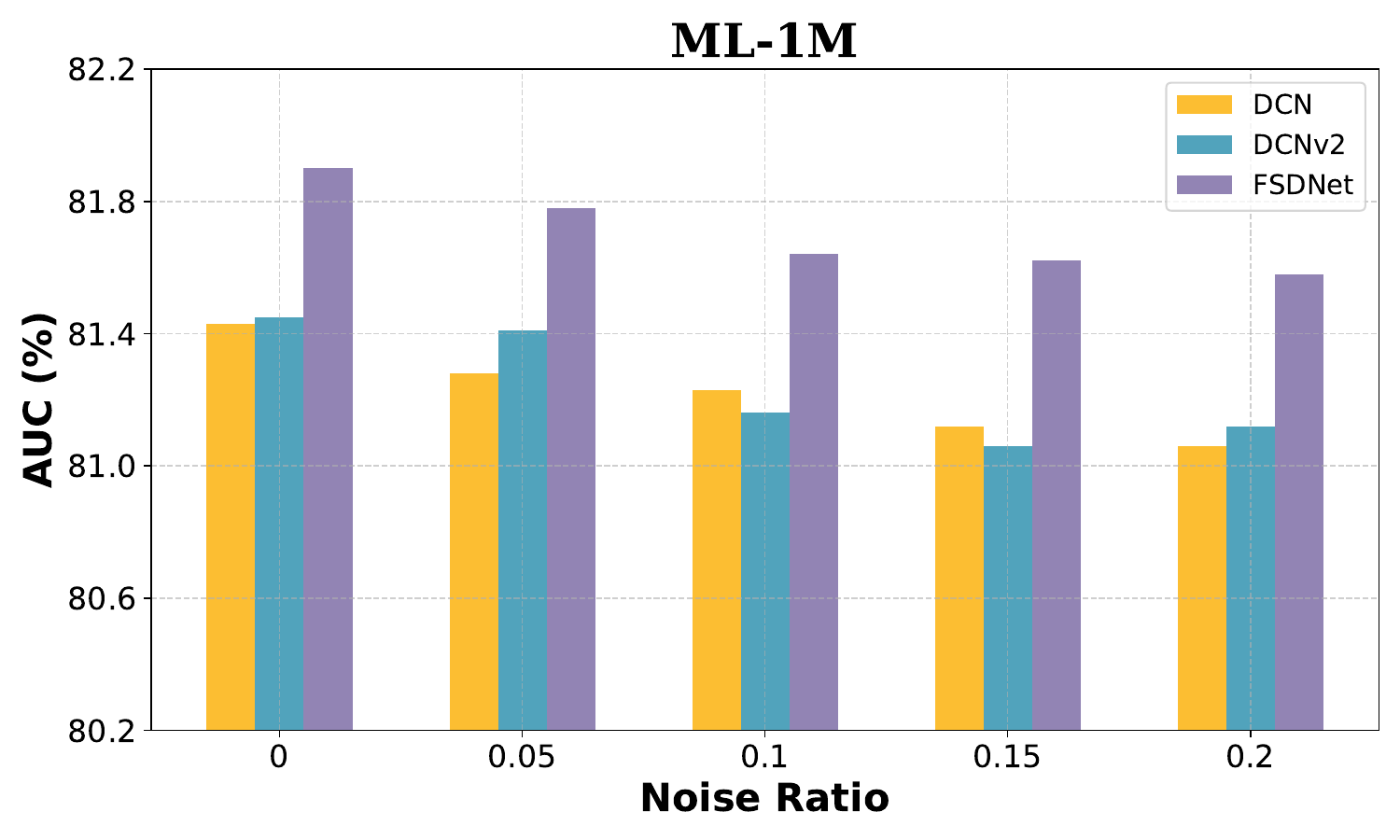}
    }
    \captionsetup{justification=raggedright}
    \caption{AUC comparison of DCN, DCNv2, and FSDNet under varying noise proportions.}
    \label{Noise}
    \vspace{-1em}
\end{figure}

\section{Related Work}
\subsection{Deep CTR Prediction}
Most of the existing deep CTR prediction models can be categorized into two types: (1) Models based on user behavior sequences \cite{2019DIEN,2022MISS,2021Coreinterest,2021sequence}, which predict the click-through probability of current recommended content by analyzing the user's behavioral sequence over a given period in the past. (2) Models based on feature interactions \cite{2016widedeep,2017DCN,2019autoint,2024Multiscenario}, which are more generalized and focus on mining the relationships between different features. These features can include user profiles, item attributes, context, etc. Since FSDNet belongs to the latter category, we briefly summarize the related works based on feature interactions. Some early methods were often constrained to modeling low or fixed-order feature interactions.  As an example, models based on FM \cite{2010FM,2016FFM} assign an embedding vector to each feature and use operations such as inner products to efficiently model second-order feature combinations. The emergence of deep learning techniques  has led to significant progress in modeling high-order implicit feature interactions.
 Given the performance limitations of MLP \cite{neuralvsmf}, many studies have integrated both explicit and implicit components. These approaches are typically divided into stacked structure \cite{2016PNN,2019FIGNN,2019OENN,2021masknet} and parallel structure \cite{2016widedeep,2017DCN,2018xdeepfm,2021dcnv2,2019autoint}, depending on how these components are combined.
\begin{itemize}
\item \textbf{Stacked Structure.} In stacked structure models, explicit feature interactions are first constructed after the embedding layer, and its output is then fed into the implicit component (usually a neural network) to further extract high-order feature interactions. Some models utilize various types of product operations (e.g., inner product, outer product, Hadamard product) to model explicit feature interactions, such as PNN \cite{2016PNN},  OENN \cite{2019OENN},and ONN \cite{2020ONN}. Some use self-attention mechanisms \cite{2017AFM} or graph neural networks \cite{2019FIGNN} to operate.
\item \textbf{Parallel Structure.} The parallel structure in CTR models typically involves modeling both explicit and implicit feature interactions in parallel using the original embeddings, and then fusing the feature information in the final output layer. By processing explicit and implicit feature interactions simultaneously, this architecture facilitates mutual reinforcement between the two.
Representative structures include Wide \& Deep \cite{2016widedeep}, DeepFM \cite{2017DeepFM}, DCN \cite{2017DCN}, xDeepFM \cite{2018xdeepfm}, DCN-v2 \cite{2021dcnv2}, and AutoInt+ \cite{2019autoint}.  Implicit feature interactions are usually captured by DNN, where the neurons in each layer learn different combinations of input features and their nonlinear relationships. Through layer-by-layer transformation, DNN can capture high-order interactions between input features.  In contrast, the methods for modeling explicit feature interactions vary significantly across different models. Wide \& Deep uses linear combinations to capture explicit feature combinations, while DeepFM leverages FM to adaptively learn pairwise feature interactions. xDeepFM introduces a CIN structure that performs layer-wise Hadamard product and convolutional operations to model high-order feature interactions. DCN and DCN-v2 propose two different types of cross networks that conduct bounded-degree feature interactions at each layer. AutoInt+ utilizes a multi-head self-attention network to capture feature interactions across different orders.
\end{itemize}

Parallel structural models have received considerable attention due to their excellent modeling and deconstruction capabilities \cite{2017DCN,2021EDCN}. However, current research mainly focuses on constructing more complex explicit feature components to boost performance, rather than improve the structure itself. 
 To address this issue, our work seeks to establish connections between explicit and implicit feature interactions at each layer, promoting the fusion of interaction information and enhancing the model’s learning capacity.

\subsection{Knowledge Distillation}  As a popular model compression approach in deep learning, knowledge distillation  \cite{2015KD,2020KDCTR} has found 
 applications across various  fields, such as image classification, speech recognition, and recommender systems.
 The core idea is to train a smaller model (the student model) to mimic the behavior of a larger and more complex model (the teacher model), thus significantly reducing the model's parameter size and computational cost while maintaining high accuracy. The traditional knowledge distillation methods \cite{2015KD,2021knowledgesurvey} treat the logit values from the teacher model as transferable knowledge.
  Subsequently, some works aimed to transfer knowledge from feature representations in the intermediate layers and final layer of deep neural networks. For instance,  \cite{2014fitnets} directly extracts semantic information from features, while  \cite{2019activationboundaries} transfers knowledge by passing the activation boundary information generated  by hidden neurons to the student network. 
Additionally, relation-based methods take into account the inter-layer relations within the model and the interconnections between samples. For example, FSP \cite{2017FSP} captures the flow between layers by calculating the inner product of features from different layers, thereby enabling more effective knowledge transfer.   \cite{2019instancedistillation} constructs an instance relationship graph to extract three types of knowledge from deep neural networks: instance features, instance relationships, and cross-layer feature space transformations. Self-distillation \cite{2019Beyourownteacher,2021knowledgesurvey} is a special type of knowledge distillation method.  Its advantage is that it does not rely on an additional teacher model, but uses the knowledge from the deepest layer to direct the learning of shallower layers in the same neural network. We attempt to apply self-distillation to the CTR task, fully leveraging the model’s potential to achieve more accurate predictions while introducing only a small number of  parameters.

\section{Conclusion}
In this paper, we propose a novel CTR prediction framework, FSDNet, which addresses several limitations in existing CTR models,  namely, insufficient information sharing, low transfer efficiency in knowledge distillation, and excessive noise in feature interactions. FSDNet establishes hierarchical connections between explicit and implicit feature interactions, enabling information sharing among different components within the parallel structure. Additionally, it leverages self-distillation to facilitate knowledge transfer within the model, enhancing learning in the shallow layers.
Experimental results demonstrate that FSDNet significantly improves CTR prediction performance across multiple benchmark datasets, exhibiting strong generalization ability and robustness. Moreover, as a lightweight module, the fusion self-distillation module can be seamlessly integrated into various parallel CTR models, effectively boosting overall performance. These contributions underscore the potential of FSDNet for broad application in real-world CTR systems.

\bibliographystyle{ACM-Reference-Format}
\bibliography{sample-base}

\appendix

\end{document}